\documentclass[times,twocolumn,final]{elsarticle}

\usepackage{framed,multirow}
\usepackage{tabularx}

\usepackage{amssymb}
\usepackage{latexsym}

\usepackage{url}
\usepackage{xcolor}
\usepackage{hyperref}
\usepackage{amsmath}
\usepackage{colortbl}
\usepackage{float}
\usepackage{threeparttable}
\usepackage{booktabs}
\usepackage{pifont}
\usepackage{bm}
\usepackage{pifont}

\usepackage{microtype}
\usepackage[modulo, switch]{lineno}

\definecolor{MyDarkGreen}{rgb}{0.0, 0.65, 0.0}
\newcommand{\RNum}[1]{\uppercase\expandafter{\romannumeral #1\relax}}

\definecolor{shadecolor}{rgb}{0.9, 0.9, 0.9}
\definecolor{aero}{rgb}{0.78515625, 1.0, 0.898039215686275}
\definecolor{faintGreen}{rgb}{0.8, 1.0, 0.8}

\definecolor{newcolor}{rgb}{.8,.349,.1}

\begin{document}



\title{MSRepaint: Multiple Sclerosis Repaint with Conditional Denoising Diffusion Implicit Model for Bidirectional Lesion Filling and Synthesis}

\author[1]{Jinwei Zhang\thanks{Corresponding author: jwzhang@jhu.edu}}
\author[2]{Lianrui Zuo}
\author[2]{Yihao Liu}
\author[3]{Hang Zhang}
\author[4]{Samuel W. Remedios}
\author[2]{Bennett A. Landman}
\author[5]{Peter~A.~Calabresi}
\author[5]{Shiv~Saidha}
\author[5]{Scott D. Newsome}
\author[1,6]{Dzung L. Pham}
\author[1,4]{Jerry L. Prince}
\author[5]{Ellen M. Mowry}
\author[1]{Aaron Carass}
 

\address[1]{Image Analysis and Communications Laboratory, Department of Electrical and Computer Engineering, Johns Hopkins University, Baltimore, MD 21218, USA}
\address[2]{Department of Electrical and Computer Engineering, Vanderbilt University, Nashville, TN 37215, USA}
\address[3]{Department of Electrical and Computer Engineering, Cornell University, Ithaca, NY 14853, USA}
\address[4]{Department of Computer Science, Johns Hopkins University, Baltimore, MD 21218, USA}
\address[5]{Department of Neurology, Johns Hopkins School of Medicine, Baltimore, MD 21287, USA}
\address[6]{Department of Radiology and Bioengineering, Uniformed Services University of the Health Sciences, Bethesda, MD, 20814, USA}


\begin{abstract}
In multiple sclerosis (MS), lesions interfere with automated magnetic resonance imaging (MRI) analyses such as brain parcellation and deformable registration, while lesion segmentation models are hindered by the limited availability of annotated training data. 
To address both issues, we propose MSRepaint, a unified diffusion-based generative model for bidirectional lesion filling and synthesis that restores anatomical continuity for downstream analyses and augments segmentation through realistic data generation.
MSRepaint conditions on spatial lesion masks for voxel-level control, incorporates contrast dropout to handle missing inputs, integrates a repainting mechanism to preserve surrounding anatomy during lesion filling and synthesis, and employs a multi-view DDIM inversion and fusion pipeline for 3D consistency with fast inference.
Extensive evaluations demonstrate the effectiveness of MSRepaint across multiple tasks. 
For lesion filling, we evaluate both the accuracy within the filled regions and the impact on downstream tasks including brain parcellation and deformable registration. 
MSRepaint outperforms the traditional lesion filling methods FSL and NiftySeg, and achieves accuracy on par with FastSurfer-LIT, a recent diffusion model-based inpainting method, while offering over 20× faster inference.
For lesion synthesis, state-of-the-art MS lesion segmentation models trained on MSRepaint-synthesized data outperform those trained on CarveMix-synthesized data or real ISBI challenge training data across multiple benchmarks, including the MICCAI 2016 and UMCL datasets. 
Additionally, we demonstrate that MSRepaint’s unified bidirectional filling and synthesis capability, with full spatial control over lesion appearance, enables high-fidelity simulation of lesion evolution in longitudinal MS progression.
Our code is available at \url{https://gitlab.com/IACL/UponAcceptance}.
\end{abstract}

\begin{keyword}
Diffusion models\sep multiple sclerosis\sep lesion filling\sep lesion synthesis\sep image inpainting
\end{keyword}

\maketitle


\section{Introduction}
\label{sec:intro}
Multiple sclerosis~(MS) is a chronic disease of the central nervous system characterized by inflammation, demyelination, and axonal damage~\citep{haider2016topograpy}.
Magnetic resonance imaging~(MRI) is widely used to detect and monitor MS lesions~\citep{bakshi2008mri}.
Automated MRI processing pipelines have significantly advanced MS imaging analysis, with key processing steps such as lesion segmentation and brain parcellation.
These steps are essential in large-scale studies of people with MS~(PwMS), supporting both clinical research and biomarker discovery.

Despite advancements in automated MRI processing pipelines, key processing steps still face major challenges due to lesion-related issues. 
First, the presence of lesions can degrade the performance of downstream algorithms, such as brain parcellation~\citep{guo2019repeatability}, cortical thickness estimation and reconstruction~\citep{magon2014white, shiee2014hbm}, and deformable image registration~\citep{sdika2009nonrigid}.
Second, the development of robust lesion segmentation models is hindered by the limited availability of high-quality training data, as manual lesion delineation is time-consuming and prone to variability across raters and imaging protocols~\citep{carass2017ni, carass2024nir}.

To mitigate these challenges, two complementary strategies have been developed for white matter T2 hyperintense lesions: \textit{lesion filling}, which inpaints lesions to generate anatomically plausible lesion-free images, and \textit{lesion synthesis}, which introduces synthetic lesions into healthy scans for training data augmentation. 
In the remainder of the introduction, we first review existing methods for lesion filling and lesion synthesis, then highlight key challenges and gaps in the current literature, and finally outline our proposed method to address these limitations.

\subsection{Lesion Filling}
Lesion filling (or lesion inpainting) is a preprocessing step that replaces voxel intensities within MS lesions with values resembling surrounding normal-appearing tissue, thereby restoring anatomical continuity. 
This helps improve downstream tasks by reducing lesion-induced misclassification in segmentation and misalignment in registration.

Early lesion filling methods assumed lesions occur in white matter~(WM) and replaced lesion voxels with intensities sampled from nearby or global WM.
For example, \citet{sdika2009nonrigid} used Gaussian-weighted local/global WM intensities, while \citet{chard2010reducing} proposed LEAP to sample from normal-appearing WM~(NAWM).
\citet{battaglini2012evaluating} implemented a similar approach in FSL~\citep{jenkinson2012fsl}.
These methods improved registration~\citep{sdika2009nonrigid} and volumetric analysis~\citep{chard2010reducing, jenkinson2012fsl}, but relied on prior tissue segmentation, which can be unreliable in MS due to lesion segmentation errors or proximity to ventricles and deep gray matter.

To address these limitations, patch-based lesion filling approaches were developed.
\citet{guizard2015non} proposed a non-local means~(NLM) inpainting method that progressively fills lesions from the edge using weighted averages of similar patches, without requiring NAWM masks.
\citet{prados2016multi} extended this to multicontrast, longitudinal data, improving upon prior NLM methods.

More recently, convolutional neural network~(CNN) based methods have been developed for lesion filling.
\citet{clerigues2023minimizing} trained a 3D patch-wise U-Net for joint tissue segmentation and lesion inpainting, effectively reducing gray matter~(GM)/WM volume estimation errors.
\citet{iglesias2023synthsr} introduced SynthSR, a CNN that standardizes heterogeneous scans into 1~mm isotropic T1w images for FreeSurfer~\citep{fischl2012freesurfer}, implicitly performing lesion inpainting by learning to ignore outlier intensities.
\citet{tang2021lg} proposed LG-Net, a U-Net with gated convolutions that adaptively fill lesion regions using learned feature masks, demonstrating robustness to imperfect lesion annotations.
\citet{zhang2020robust} proposed an edge-guided network that incorporates Canny-derived edge maps to guide lesion filling and preserve anatomical structure.

Generative models have also been applied to lesion filling.
\citet{pollak2025fastsurfer} introduced FastSurfer-LIT, a lesion filling method based on denoising diffusion probabilistic models~(DDPMs)\citep{ho2020denoising} and inspired by the RePaint approach for natural images~\citep{lugmayr2022repaint}.
FastSurfer-LIT is integrated into the FastSurfer pipeline~\citep{henschel2020fastsurfer} and supports robust surface reconstruction and volumetric analysis in brains affected by lesions or cavities.

\subsection{Lesion Synthesis}
While lesion filling removes lesions to restore normal anatomy, lesion synthesis does the opposite—generating realistic lesions in lesion-free MRI.
It serves two main purposes: (1)~augmenting training data for lesion segmentation models, and (2)~creating test cases with lesion-free ground truth for evaluating lesion filling and other methods.

Early synthesis approaches inserted artificial lesions into healthy scans using geometric models or real lesion masks.
\citet{chard2010reducing} manually added spheroid lesions using ImageJ\footnote{\url{https://imagej.net/ij/}}, controlling lesion load, intensity, and anatomical distribution to mimic realistic MS pathology.
\citet{battaglini2012evaluating} overlaid real MS lesion masks on healthy brains and filled them with Gaussian-sampled intensities mimicking cerebrospinal fluid~(CSF), GM, or tissue boundaries.
\citet{sdika2009nonrigid} and \citet{guizard2015non} adapted the method from \citet{brett2001spatial}, applying voxel-wise intensity ratios from patient lesions to registered healthy WM.

For data augmentation, copy-paste strategies have gained popularity due to their simplicity and ability to preserve realistic lesion appearance.
Inspired by work in computer vision~\citep{ghiasi2021simple}, these methods copy real lesions from one image and paste them into another.
CarveMix\citep{zhang2023carvemix} is a representative copy-paste method that carves variable-sized lesion regions while preserving local anatomical context, harmonizing intensities across scans, and modeling mass effect for large lesions through deformation.
Similar methods such as TumorCP~\citep{yang2021tumorcp} and Soft Poisson Blending~\citep{huo2025self} enhance realism by incorporating smoothing or blending techniques during the lesion pasting process.

CNN-based lesion synthesis models have also been explored to generate realistic lesions for data augmentation.
\citet{salem2019multiple} trained an encoder–decoder CNN to synthesize MS lesions by using lesion-filled pseudo-healthy images as inputs and the corresponding original lesioned images as targets.
They demonstrated the effectiveness of their method for training data augmentation on the ISBI 2015 lesion segmentation challenge~\citep{carass2017dib}.

Deep generative models have further advanced lesion synthesis. \citet{basaran2022subject} employed a generative adversarial network~(GAN)\citep{goodfellow2014generative} with attention mechanisms and a lesion-aware discriminator to synthesize MS lesions on healthy brain MRIs and generate pseudo-healthy counterparts from pathological scans.
\citet{zhang2024lefusion} proposed LeFusion, a lesion-focused diffusion model for controllable lesion synthesis, though it was not assessed on MS lesions.
\citet{mathur2025long} proposed a conditional diffusion model to generate anatomically plausible follow-up MRI scans with corresponding new MS lesion labels, aiming to augment training data for longitudinal lesion assessment.

\subsection{Challenges and Gaps}

Learning-based methods, including CNNs and generative models, have made significant improvements over early lesion filling and synthesis methods in terms of filling accuracy and synthesis realism.
However, we identified four key challenges and gaps faced by learning-based methods in the literature:


First, although DDPM-based methods~\citep{pollak2025fastsurfer, zhang2024lefusion, mathur2025long} have outperformed GAN-based methods~\citep{basaran2022subject} in generation quality and training stability, they remain highly resource-intensive in training.
For instance, \citet{zhang2024lefusion} and \citet{mathur2025long} used multiple high-end A100 GPUs for training, while \citet{pollak2025fastsurfer} trained their model on thousands of 3D MRI volumes.
These substantial computational and data requirements pose barriers for researchers seeking to reproduce or adapt such models in low-resource or small-data settings.

Second, achieving inter-slice consistency and efficient inference time with DDPMs for 3D MRI synthesis remains challenging.
\citet{pollak2025fastsurfer} employed a multi-view strategy using three 2D DDPMs (axial, sagittal, and coronal) to ensure 3D consistency; however, in our testing on a standard GPU (NVIDIA Quadro RTX 5000), this approach required approximately one hour of runtime for processing a single T1w 3D brain MRI volume.
\citet{zhang2024lefusion} trained a full 3D DDPM on small patches (e.g., $64 \times 64 \times 32$), achieving inherent 3D consistency but at the cost of high computational demand, making it impractical for full-resolution brain MRI (e.g., $256 \times 256 \times 256$).

Third, most existing models are designed for either a single MRI contrast or a fixed combination of contrasts, limiting their flexibility.
Models designed on a single input contrast (e.g., FLAIR only in \citet{basaran2022subject}) cannot be used in multicontrast settings, where physiologically consistent lesion appearances and complementary information across sequences such as T1w, T2w, and FLAIR are essential.
Meanwhile, models designed on fixed contrast combinations (e.g., T1w and FLAIR in \citet{salem2019multiple}) cannot accommodate missing or incomplete contrasts, which is a common scenario in clinical practice and a key requirement for real-world deployment.

Finally, most existing methods support either lesion filling or lesion synthesis, but not both, which limits their ability to simulate the full range of lesion dynamics (growth, shrinkage, appearance, disappearance) in MS follow-up scans.
To the best of our knowledge, \citet{basaran2022subject} is the only bidirectional model for both tasks, but it is fundamentally a GAN-based approach which suffers from training instability and mode collapse~\citep{mescheder2018training}, and was trained only on central axial FLAIR slices.
For follow-up MRI simulation, \citet{mathur2025long} developed a diffusion model that generates new lesions at follow-up given a baseline FLAIR MRI and the corresponding lesion mask, but it is limited to 2D FLAIR slices and cannot model other lesion dynamics.

\subsection{Our Contributions}
Inspired by RePaint~\citep{lugmayr2022repaint} for 2D natural image inpainting, our proposed method, \textbf{MSRepaint}, makes the following contributions, each addressing one of the aforementioned challenges and gaps in order:
\begin{itemize}
    
    \item \textbf{Resource-efficient training:} MSRepaint achieves high-quality results trained with limited data and computational resources.

    \item \textbf{Fast and inter-slice consistent inference:} MSRepaint employs a multi-view sequential DDIM~\citep{song2021iclr} inversion and fusion pipeline that ensures efficient and 3D consistent inference.
    
    \item \textbf{Multicontrast consistency with dropout robustness:} MSRepaint ensures physiologically consistent lesion synthesis across T1w, T2w, and FLAIR, and remains robust to missing or partial contrast inputs.

    \item \textbf{Unified lesion filling and synthesis:} MSRepaint unifies filling and synthesis within a single framework, providing voxel-wise control of lesion appearance and enabling high-fidelity simulation of longitudinal lesion evolution.
\end{itemize}

This work extends our conference paper~\citep{zhang2025bi}, where we initially introduced the core idea of MSRepaint with preliminary results. 
This work systematically presents the MSRepaint methodology along with several key extensions, including multi-view sequential DDIM inference and fusion, multicontrast dropout, and comprehensive evaluations of its lesion filling and synthesis capabilities.
We benchmark MSRepaint against established methods across multiple experiments and demonstrate its superior performance.

\section{Method}
\label{s:method}

\begin{figure*}[!t]
    \centering
    \includegraphics[width=2.0\columnwidth]{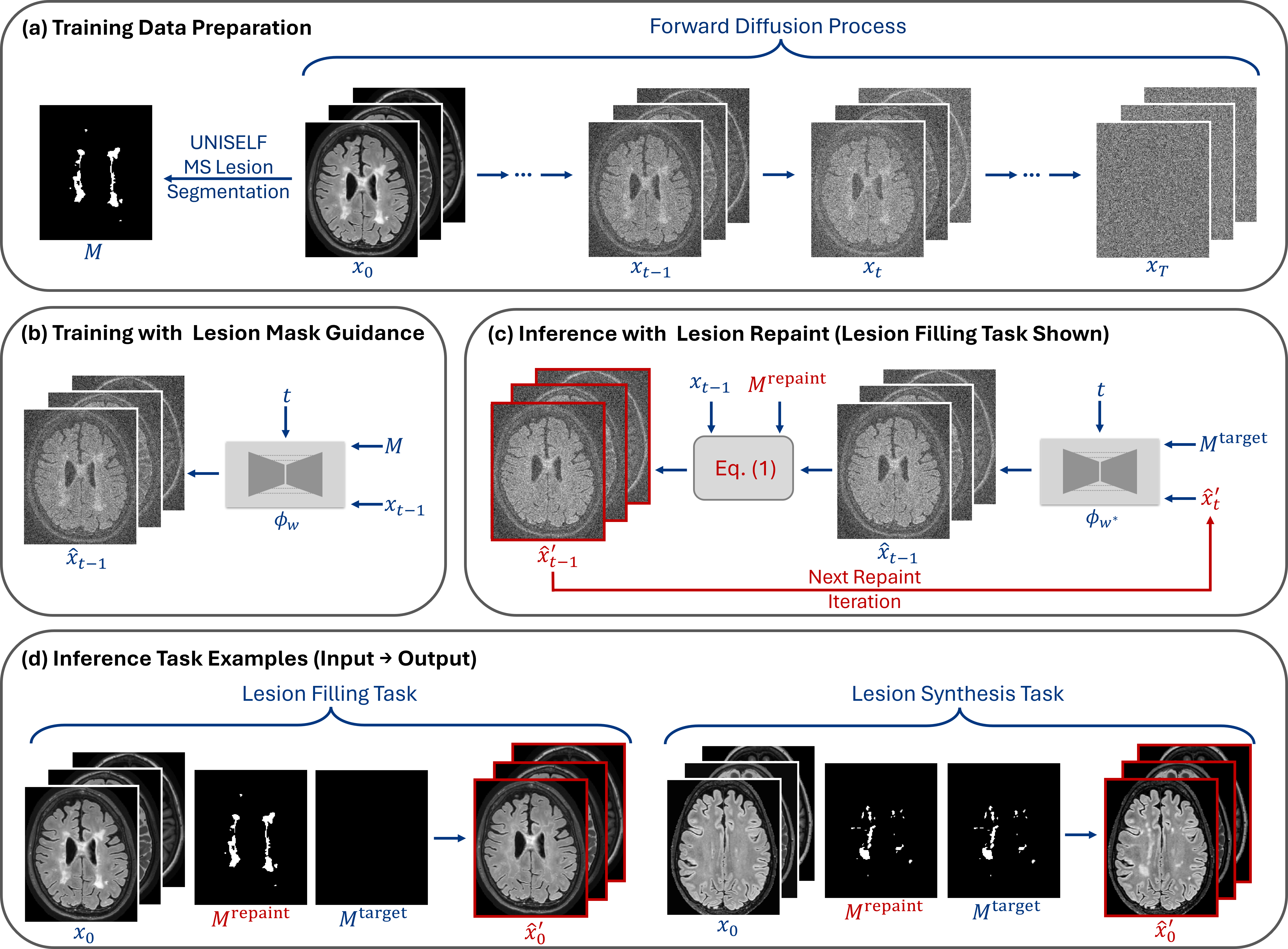}
    \caption{Overview of MSRepaint training and inference.
    (a) Training data: A 2D multicontrast image $x_0$ undergoes forward diffusion, generating noisy images $x_{t-1}$, $x_t$, and the final degraded image $x_T$, paired with its lesion mask $M$. 
    (b) Training: The denoiser $\phi_w$ is trained to estimate $\hat{x}_{t-1}$ from $x_t$ with guidance from mask $M$.
    (c) Inference: The trained denoiser $\phi_{w^*}$ generates outputs guided by $M^{\text{target}}$, while the repaint mask $M^{\text{repaint}}$ restricts updates to its region, preserving surrounding tissue. 
    A repainting mechanism iteratively refines update boundaries to ensure seamless integration with the surrounding tissue.
    (d) Example tasks: Lesion filling sets $M^{\text{target}}=0$ to replace lesions with healthy tissue, while lesion synthesis sets $M^{\text{target}}=M^{\text{repaint}}$ to generate new lesion patterns.
    }
    \label{fig:figure1}
\end{figure*}

We propose MSRepaint~(Multiple Sclerosis Repaint) as a unified diffusion model for bidirectional MS lesion filling and synthesis.
In MSRepaint, lesion filling and synthesis can both be formalized as image inpainting problems, where the objective is to modify voxel intensities within a given region of interest~(ROI).
The two primary tasks of MSRepaint in this context are: (1)~determining, on a per-voxel basis, where to perform lesion filling or synthesis, and (2)~replacing lesion voxels with healthy-appearing tissue for filling, or generating realistic lesion patterns in normal-appearing regions for synthesis.

\subsection{Training Data Preparation}
\label{subsec:training_data_prep}
To prepare training data for MSRepaint, we used a multicontrast MRI dataset of PwMS, together with their corresponding MS lesion segmentation derived by UNISELF~\citep{zhang2024towards}.
From this dataset, 2D multicontrast MR images in axial, coronal, and sagittal views were extracted. 
Given such a 2D image $x_0$ (formed by concatenating T1w, T2w, and FLAIR contrasts along the channel dimension), a forward diffusion process ${x_{1:T}}$ was applied by gradually adding Gaussian noise to $x_0$ according to a predefined mean and variance schedule over timesteps $1$ to $T$~\citep{ho2020denoising}.
An example 2D image $x_0$ with its corresponding lesion mask $M$ is shown in Fig.~\ref{fig:figure1}(a), together with noisy images at intermediate timesteps $t-1$ and $t$, and the final image $x_T$ at the end of the forward diffusion process. 

\subsection{Training with Lesion Mask Guidance}
\label{subsec:training_w_lesions}
To train the reverse (denoising) process, a time-embedded U-Net-based denoiser~\citep{ronneberger2015u}, denoted as $\phi_w$ with learnable weights $w$, is used to predict the added noise $\epsilon_t$ at each timestep $t$.
Specifically, $\phi_w$ takes as input the noisy image $x_t$ and its corresponding mask $M$, and outputs a noise estimate $\hat{\epsilon}_t = \phi_w(M, x_{t-1}; t)$. This estimate is used to estimate either the clean image $\hat{x}_0$ or the image at the previous timestep $\hat{x}_{t-1}$, as shown in Fig.~\ref{fig:figure1}(b), via the reparameterization process defined by the forward diffusion schedule~\citep{ho2020denoising}.
The training loss $|| \epsilon_t - \hat{\epsilon}_t ||^2$ is modulated based on the values of the conditioning mask $M$.
For lesion-free inputs (2D images for which there are no lesions), the loss is computed over the entire image to capture the distribution of healthy tissue. 
For inputs containing lesions, the loss is upweighted by a factor within the lesion regions, encouraging $\phi_w$ to focus on synthesizing realistic lesion patterns guided by the input lesion mask.

\subsection{Training with Contrast Dropout}
\begin{figure*}[!t]
    \centering
    \includegraphics[width=2.0\columnwidth]{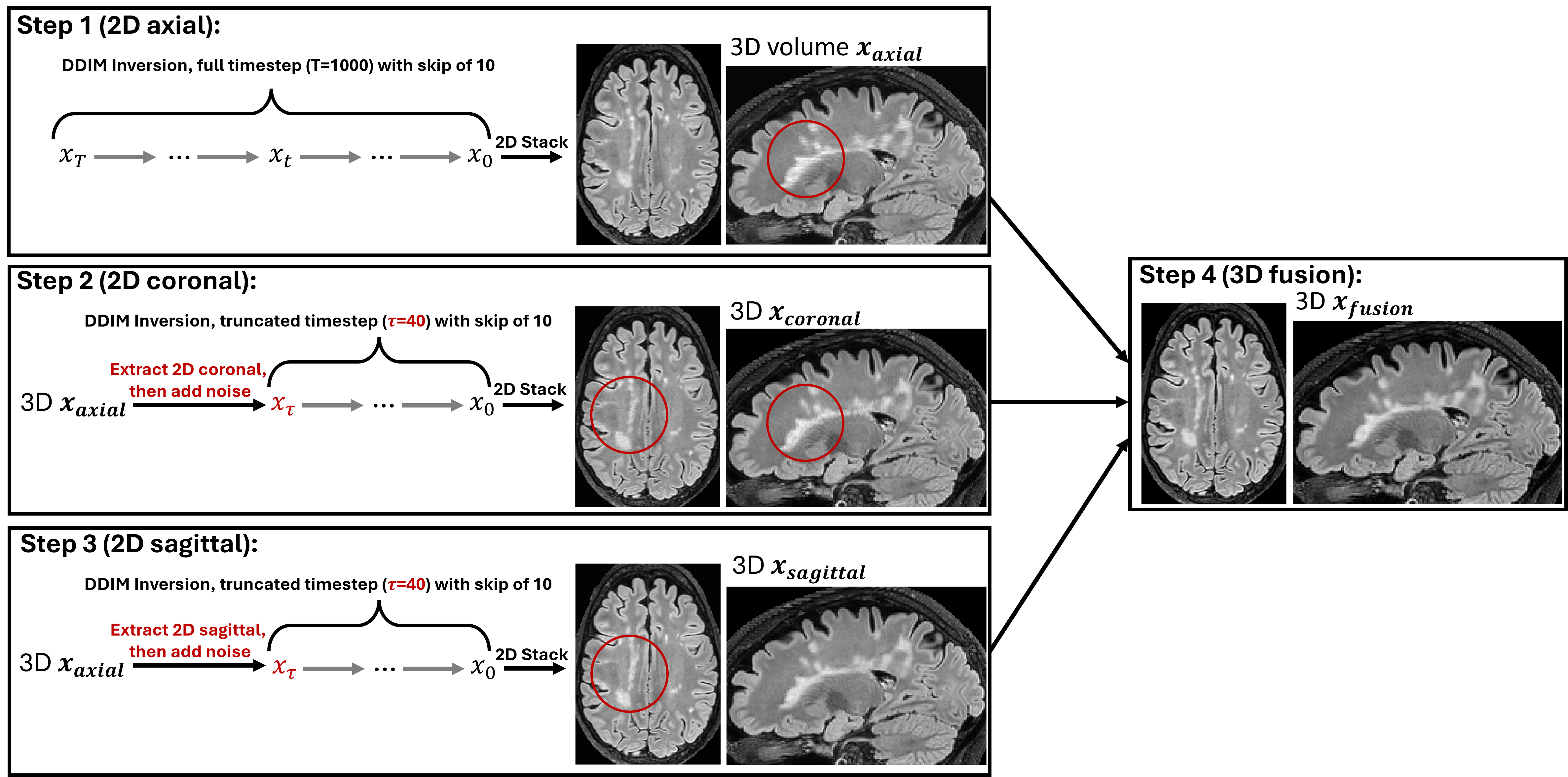}
    \caption{Overview of the multi-view DDIM inversion and fusion pipeline.
    Step 1: A full DDIM inversion is performed on 2D axial slices to generate a 3D image $x_{\text{axial}}$.
    Step 2: The resulting $x_{\text{axial}}$ is reoriented into the coronal view, and a truncated DDIM inversion is applied to 2D coronal slices with added noise, yielding $x_{\text{coronal}}$.
    Step 3: The same process is repeated for the sagittal view to obtain $x_{\text{sagittal}}$.
    Step 4: The three outputs, $x_{\text{axial}}$, $x_{\text{coronal}}$, and $x_{\text{sagittal}}$, are fused to produce the final 3D image $x_{\text{fusion}}$ with enhanced inter-slice consistency.
    Red circles highlight synthetic lesion regions that exhibit inter-slice inconsistency in each view, which is resolved after the final fusion step.
}
    \label{fig:figure2}
\end{figure*}

\label{subsec:training_w_cd}
A contrast dropout training strategy~\citep{feng2019self} is employed, where for each training batch, a random subset of the full input contrasts is dropped by zeroing out the corresponding channels. 
This encourages the model to be robust to missing contrasts and improves generalization to clinical scenarios with incomplete acquisitions.

\subsection{Inference with Lesion Repaint}
 
Inference is guided by two complementary masks: $M^{\text{target}}$ and $M^{\text{repaint}}$.
The target mask $M^{\text{target}}$, provided as part of input to the trained denoiser $\phi_{w^*}$ during inference, specifies the desired lesion configuration to be generated by $\phi_{w^*}$.
The repaint mask $M^{\text{repaint}}$ defines the ROI to be updated. 
Specifically, at each timestep $t$, voxels outside $M^{\text{repaint}}$ are preserved from the true noisy image $x_{t-1}$, whereas those inside are replaced with newly synthesized content by $\phi_{w^*}$. 
This is achieved by updating the denoised estimate $\hat{x}_{t-1}$ after each reverse diffusion step through mixing it with the true noisy image $x_{t-1}$ from the forward process:
\begin{align}
\label{eq:mix}
    \hat{x}_{t-1}' = \hat{x}_{t-1} \cdot M^{\text{repaint}} + x_{t-1} \cdot (1 - M^{\text{repaint}}).
\end{align}
This ensures that only voxels within $M^{\text{repaint}}$ are modified.   

To further ensure anatomical fidelity and seamless integration with surrounding tissue, MSRepaint incorporates a repainting mechanism inspired by RePaint~\citep{lugmayr2022repaint}, as illustrated in Fig.~\ref{fig:figure1}(c), where red colored symbols and text highlight all variables and operations related to the repainting mechanism. 
After obtaining $\hat{x}_{t-1}'$, a new sample $\hat{x}_t'$ is regenerated from $\hat{x}_{t-1}'$ through one forward diffusion step and then passed back through the reverse process at timestep $t$. 
This step, denoted as the ``Next Repaint Iteration'' in Fig.~\ref{fig:figure1}(c), leads to repainting the region defined by $M^{\text{repaint}}$ to ensure seamless integration with the surrounding tissue~\citep{lugmayr2022repaint}.
This repaint iteration can be repeated multiple times, with the number of repetitions treated as a tunable hyperparameter.

This dual-mask strategy enables both lesion filling and lesion synthesis tasks within a unified framework. 
For lesion filling, $M^{\text{target}}$ is set to all zeros, directing the model to replace lesion voxels in $M^{\text{repaint}}$ with healthy-appearing tissue. 
For lesion synthesis, $M^{\text{target}}$ is set equal to $M^{\text{repaint}}$, instructing the model to generate lesion-like patterns in the designated healthy regions. 
Representative input–output examples of both tasks are shown in Fig.~\ref{fig:figure1}(d).

\subsection{Inference with DDIM Acceleration}
To accelerate the inference process, we apply a DDIM~\citep{song2021iclr} with a sub-sequence $[\tau_1, \cdots, \tau_S]$ of $[1, \cdots, T]$ for reverse sampling ($S \ll T$).
Since the foward process is no longer Markovian in the DDIM, we regenerate $\hat{x}_t'$ for lesion repaint using an inpainted ${\hat{x}_0}'$ instead of ${\hat{x}_{t-1}}'$ in Eq.~\ref{eq:mix}.
Inpainting ${\hat{x}_0}$ is similar to Eq.~\ref{eq:mix} with $t=0$: ${\hat{x}_0}' = {\hat{x}_0} \cdot {M^\text{repaint}} + {x_0} \cdot ({1} - {M^\text{repaint}})$, where $\hat{x}_0$ on the right-hand side is also estimated from the reverse process at timestep $t$.

\subsection{Inference with Multi-view Inversion and Fusion}

To improve inter-slice consistency, we propose a novel multi-view DDIM inversion and fusion pipeline, as illustrated in Fig.~\ref{fig:figure2} using a lesion synthesis example.
The pipeline consists of the following steps:

Step 1: A full DDIM inversion is performed on 2D axial slices to generate an initial 3D volume, denoted as $\boldsymbol{x}_{\text{axial}}$.

Step 2: The volume $\boldsymbol{x}_{\text{axial}}$ is reoriented into the coronal view. 
Gaussian noise is then added to these 2D coronal slices, followed by a truncated DDIM inversion to recover the image in the coronal plane, yielding $\boldsymbol{x}_{\text{coronal}}$.

Step 3: The same process is applied in the sagittal view using $\boldsymbol{x}_{\text{axial}}$ as input, yielding $\boldsymbol{x}_{\text{sagittal}}$.

Step 4: The three intermediate outputs, $\boldsymbol{x}_{\text{axial}}$, $\boldsymbol{x}_{\text{coronal}}$, and $\boldsymbol{x}_{\text{sagittal}}$, are fused using a dedicated fusion network~\citep{zuo2023haca3} for lesion synthesis, or by taking the per-voxel median of their intensity values for lesion filling, to generate the final 3D image $\boldsymbol{x}_{\text{fusion}}$.

As highlighted by the red circles in Fig.~\ref{fig:figure2} using a lesion synthesis example, the fusion step corrects inter-slice inconsistencies in the synthetic lesion appearance of each view.
Overall, the proposed multi-view DDIM pipeline significantly improves lesion synthesis quality and 3D spatial consistency while preserving efficient inference runtime.

\subsection{Implementation Details}
The following training configurations were used:
\begin{itemize}
    \item The reverse process was modeled using a time-embedded U-Net backbone with $T=1000$ time steps. 
    \item A cosine noise schedule of the forward process was applied for improved image quality~\citep{nichol2021improved}.
    \item A weighting factor of 10 was applied to the lesion regions in the training loss function.
    \item The Adam optimizer was used with an initial learning rate of $3\times10^{-4}$, a batch size of 32 randomly selected from 2D slices in the three cardinal planes, and was conducted over 300 epochs.
\end{itemize}

The following inference configurations were used:
\begin{itemize}
    \item DDIM inference used a subset of the original time steps, with every 10$^{\text{th}}$ step included, resulting in 100 DDIM reverse steps in the axial view.
    \item Truncated DDIM inversion was performed with $\tau = 40$ as the starting timestep, resulting in 4 DDIM reverse steps in the sagittal and coronal views.
    \item Lesion repaint was repeated twice at each timestep.
\end{itemize}

\subsection{Large-scale Synthesis with Lesion Dictionaries}
\label{subsec:lesion_dictionary_and_synthesis}

\begin{figure}[!t]
	\centering
    \includegraphics[width = 1.0\columnwidth]{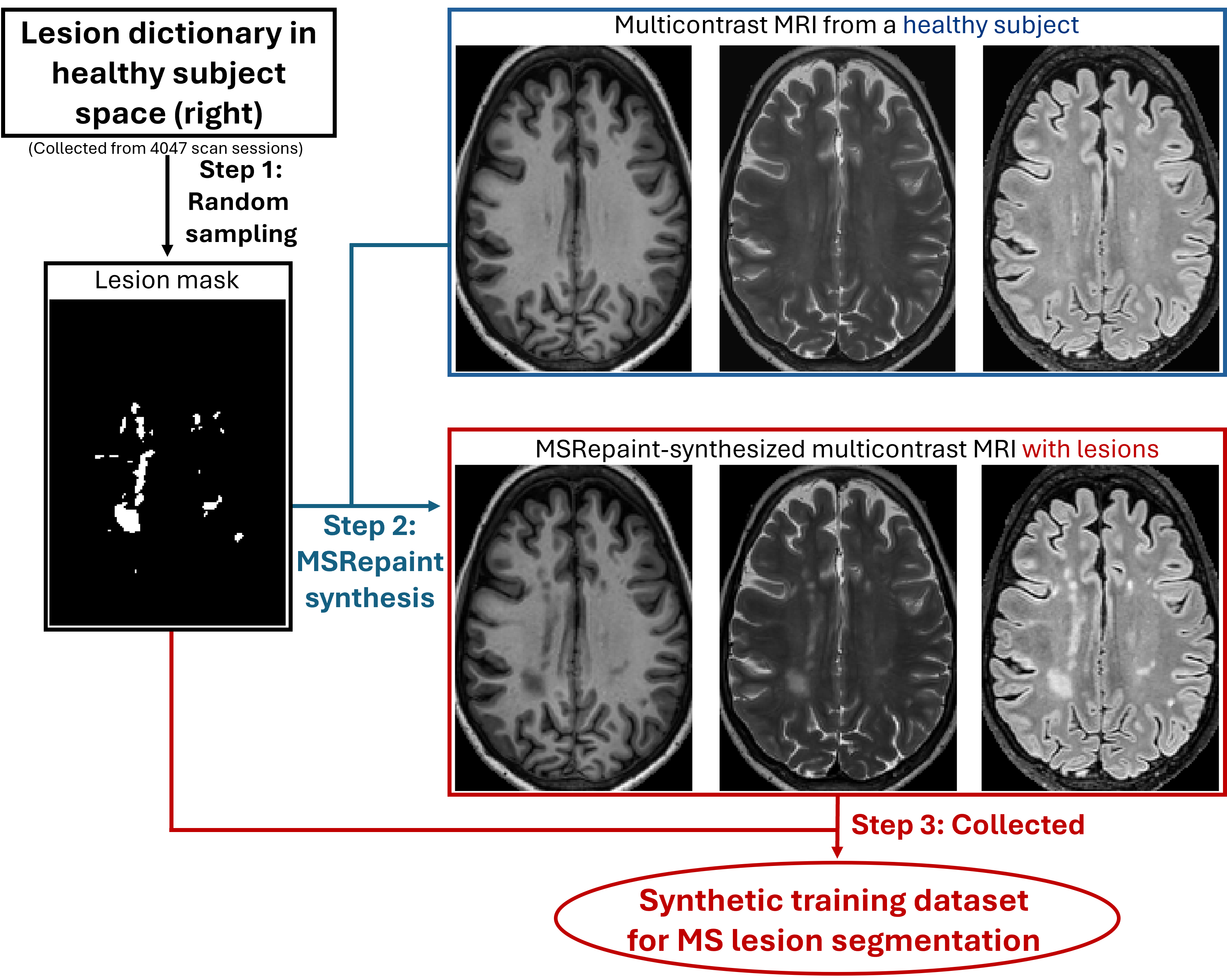}
	\caption{Pipeline for generating a synthetic training dataset for MS lesion segmentation using MSRepaint with lesion dictionaries. 
    Five lesion dictionaries were created, each aligned to the anatomical space of a different healthy subject via deformable registration. 
    Step 1: Lesion masks were randomly sampled from each dictionary. 
    Step 2: Sampled masks and corresponding multicontrast MRI scans from healthy subjects were input into the MSRepaint model to generate lesion-synthesized images. 
    Step 3: The synthesized images and their associated lesion masks were compiled to construct a synthetic training dataset.}
	\label{fig:figure9}
\end{figure}

To generate large-scale synthetic training data using MSRepaint for MS lesion segmentation, we developed a pipeline based on lesion dictionaries (see~Fig.~\ref{fig:figure9}). 
This pipeline has three stages: (1)~construction of lesion dictionaries, (2)~sampling of candidate lesion masks, and (3)~synthesis of training images using MSRepaint. 
Each stage is described below.

\subsubsection{Lesion Dictionary Construction}
\label{subsubsec:lesion_dictionary_generation}
Five lesion dictionaries were built using the Private Multisite II dataset (Dataset~\#5 in Table~\ref{tab:dataset_summary}), which includes 4,047 multicontrast sessions from 1,132 PwMS. 
Each dictionary was anchored to the anatomical space of a different lesion-free multicontrast image from a healthy subject.
An example lesion-free multicontrast image used as the basis of one dictionary is shown in the top row of Fig.~\ref{fig:figure9}.

For each dictionary, all 4,047 imaging sessions from PwMS were first lesion-filled on their T1w images and then deformably registered to the lesion-free target subject that defined the dictionary space.
The resulting deformation fields were used to warp the binary lesion masks from each PwMS session into that subject’s anatomical space.
This procedure produced five target-specific dictionaries, each consisting of thousands of lesion masks spatially aligned within the coordinate system of a single healthy subject.

\subsubsection{Candidate Lesion Mask Sampling}
\label{subsubsec:candidate_lesion_mask}
To generate one candidate lesion mask from one of the dictionaries, we employed a three-step sampling process. 
First, eight scan sessions are randomly selected from the 4,047 binary lesion masks in the dictionary.
Second, from the eight selected sessions, one-eighth of the connected lesion components was randomly chosen.
Finally, the selected components were aggregated to form a composite mask. 
This strategy introduced diversity in the number, size, and spatial distribution of the lesions while preserving anatomical plausibility with the base dictionary image. 
An example candidate mask produced by this process is shown as the output of Step~1 in Fig.~\ref{fig:figure9}.

\subsubsection{Synthetic Dataset Generation}
\label{subsubsec:synthetic_dataset_generation}

\begin{table*}[!t]
\centering
\caption{Summary of datasets used in this study.}
\label{tab:dataset_summary}
\footnotesize
\begin{tabular}{>{\centering\arraybackslash}p{0.6cm} >{\raggedright\arraybackslash}p{2.8cm} >{\raggedright\arraybackslash}p{3.0cm} >{\centering\arraybackslash}p{1.0cm} >{\raggedright\arraybackslash}p{3.3cm} >{\raggedright\arraybackslash}p{4.5cm}}
\toprule
\textbf{ID} & \textbf{Dataset} & \textbf{Usage} & \textbf{Type} & \textbf{Notes} & \textbf{Evaluation Protocol} \\
\midrule
\addlinespace
0 & Private Multisite I & MSRepaint training~(Sections~\ref{subsec:training_data_prep},~\ref{subsec:training_w_lesions} and~\ref{subsec:training_w_cd}) & Real & 30,400 axial, sagittal, and coronal slices from 76 subjects across 9 sites; T1w, T2w, and FLAIR contrasts with lesion masks predicted by UNISELF~\citep{zhang2024isbi,zhang2025uniself} & N/A \\
\hline \addlinespace
1 & Public BrainWeb~\citep{cocosco1997brainweb} & Lesion filling evaluation~(Section~\ref{subsubsec:lesion_filling_brainweb}) & Synthetic & One T1w image with severe lesions and a corresponding lesion-free T1w reference & Root mean squared error (RMSE) between the lesion-filled and lesion-free images within lesion regions \\
\hline \addlinespace
2 & Private Real Lesions & Lesion filling evaluation~(Section~\ref{subsubsec:lesion_filling_brain_vol}) & Real & 15 T1w images from PwMS with lesion masks predicted by UNISELF~\citep{zhang2024isbi, zhang2025uniself} & Brain segmentation Dice score by registering the lesion-filled image to a lesion-free target and comparing the warped segmentation to the target segmentation \\
\hline \addlinespace
3 & Private Synthetic Lesions & Lesion filling evaluation~(Section~\ref{subsubsec:lesion_filling_deform}) & Synthetic & 10 T1w images from healthy subjects with synthetic lesions copy-paste from separate MS subjects & Deformation field RMSE within lesion regions by comparing the field from the lesion-filled image to that from the corresponding lesion-free image \\
\hline \addlinespace
4 & Private Multisite II & Synthetic training dataset generation for lesion segmentation via constructed lesion dictionaries~(Section~\ref{subsec:lesion_dictionary_and_synthesis}) & Real & 4,047 multicontrast MRI sessions from 1,131 PwMS across multiple sites, with lesion masks predicted by UNISELF~\citep{zhang2024isbi, zhang2025uniself} & N/A \\
\hline \addlinespace
5 & Public ISBI 2015~\citep{carass2017longitudinal} & Synthetic training dataset generation for lesion segmentation and baseline lesion segmentation training~(Section~\ref{subsubsec:carvemix_msrepaint_synthesis}) & Real & 21 scan sessions from 5 subjects, each with lesion annotations from 2 experts & N/A \\
\hline \addlinespace
6 & Public MICCAI 2016~\citep{commowick2018objective} & Lesion synthesis evaluation~(Sections~\ref{subsubsec:lesion_seg_qualitative} and~\ref{subsubsec:lesion_seg_quantitative}) & Real & 15 subjects from 3 scanners; consensus lesion annotations from 7 experts & Dice Similarity Coefficient~(DSC), Lesion-wise True Positive Rate~(L-TPR), Lesion-wise False Positive Rate~(L-FPR), Precision, Sensitivity, Voxel-wise F1~(V-F1), and Lesion-wise F1~(L-F1) \\
\hline \addlinespace
7 & Public UMCL~\citep{lesjak2018novel} & Lesion synthesis evaluation~(Sections~\ref{subsubsec:lesion_seg_qualitative} and~\ref{subsubsec:lesion_seg_quantitative}) & Real & 30 subjects from 1 scanner; consensus lesion annotations from 3 experts & Same as Dataset~\#6 \\
\hline \addlinespace
8 & Private Longitudinal Data & Lesion evolution simulation evaluation~(Section~\ref{subsec:lesion_evo}) & Synthetic & Multicontrast MRI of one healthy subject with simulated longitudinal lesion changes using custom masks at two follow-up timepoints & Visual evaluation of lesion appearance, disappearance, growth, and shrinkage using MSRepaint over time \\
\bottomrule
\end{tabular}
\end{table*}

In this stage, each sampled lesion mask was combined with its corresponding  multicontrast lesion-free target and input into MSRepaint to synthesize new images with lesions (Step~2 in Fig.~\ref{fig:figure9}). 
Finally, the synthesized multicontrast images and their corresponding ground-truth lesion masks were collected to construct a large-scale training dataset for MS lesion segmentation (Step~3 in Fig.~\ref{fig:figure9}).

\section{Experiments and Results}
\label{s:experiments}

\subsection{Datasets and Experimental Design}
Table~\ref{tab:dataset_summary} provides a detailed overview of all datasets used in this work.  
These include: MSRepaint training dataset (\#0), lesion filling evaluation datasets (\#1--3), datasets to generate synthetic training data for lesion segmentation (\#4--5), lesion synthesis evaluation datasets (\#6--7), and lesion evolution evaluation dataset (\#8).
The following subsections present experimental results on these datasets.
Specifically:
\begin{itemize}
    \item Sections~\ref{subsubsec:lesion_filling_brainweb}–\ref{subsubsec:lesion_filling_deform} benchmark lesion filling methods on BrainWeb simulations, registration-based volumetric analysis, and deformation field accuracy.
    \item Sections~\ref{subsubsec:carvemix_msrepaint_synthesis}–\ref{subsubsec:lesion_seg_quantitative} compare lesion synthesis against CarveMix and assess the impact of different synthetic datasets on downstream lesion segmentation performance.
    \item Section~\ref{subsec:lesion_evo} evaluates MSRepaint’s ability to simulate longitudinal lesion evolution using controlled masks.
\end{itemize}

\begin{figure*}[!t]
	\centering
    \includegraphics[width=2.0\columnwidth]{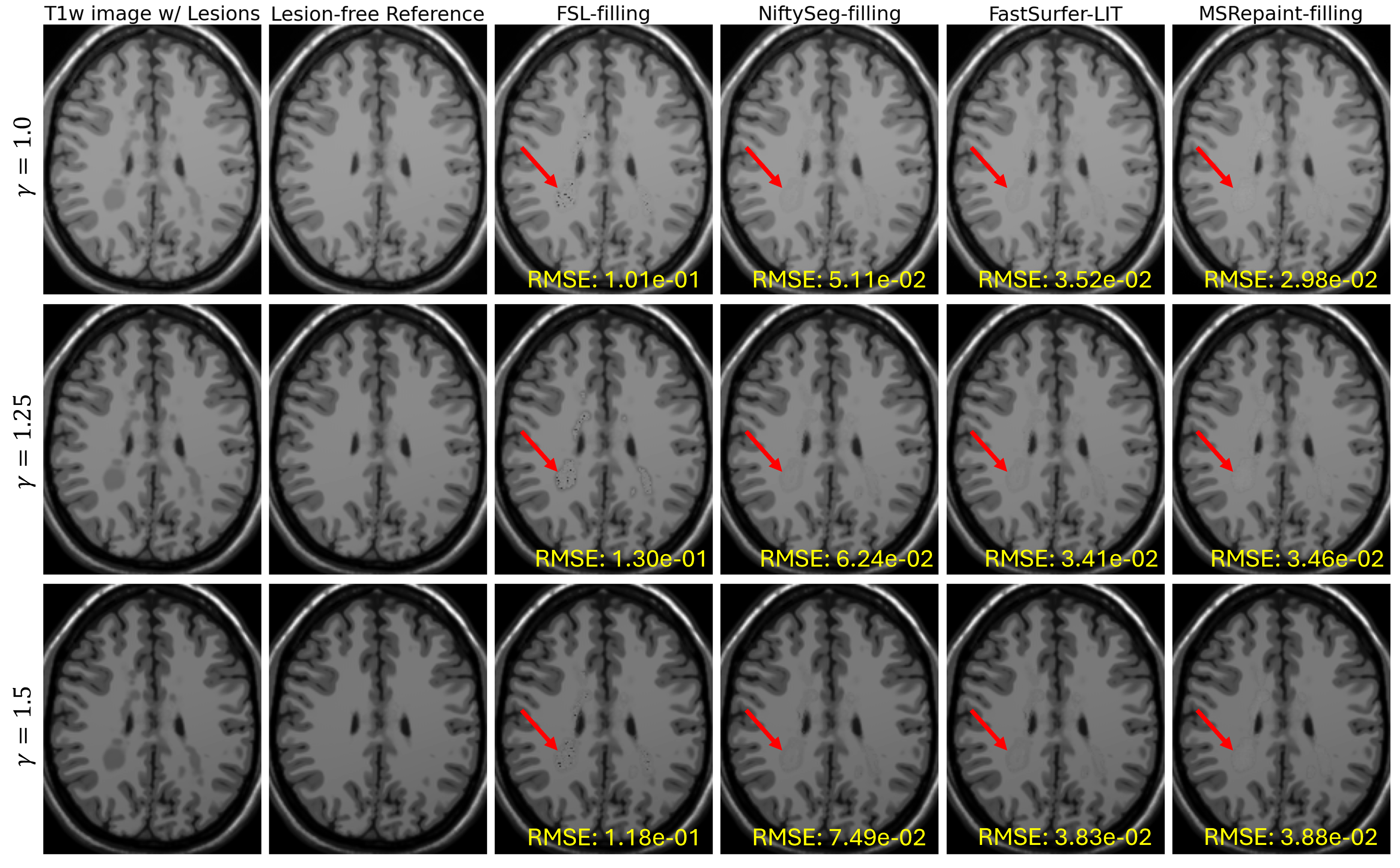}
	\caption{Lesion-filling accuracy comparison under increasing lesion-to-NAWM contrast on BrainWeb data (simulated with $\gamma = 1.0$, $1.25$, and $1.5$). 
    Root mean squared error (RMSE) was computed within the lesion mask and normalized by the intensity of the surrounding NAWM. 
    Red arrows highlight visual differences in the lesion-filled regions produced by the different lesion-filling methods.
	}
	\label{fig:figure3}
\end{figure*}

\subsection{Lesion Filling Evaluation}

\begin{figure*}[!t]
	\centering
    \includegraphics[width=2.0\columnwidth]{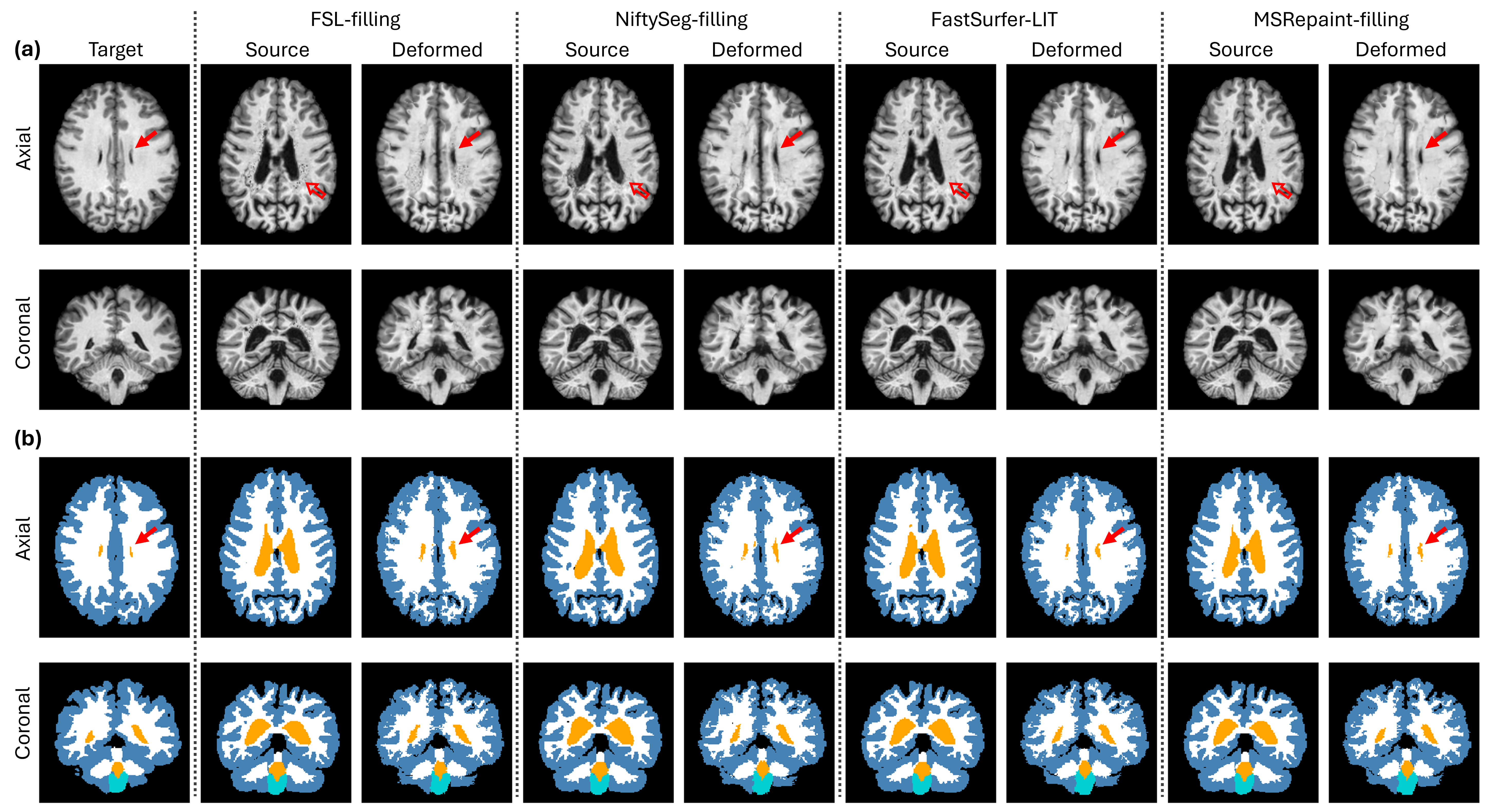}
	\caption{Comparison of lesion-filling methods on registration-based brain volume measurements using VoxelMorph.
    (a)~Lesion-filled source images and their corresponding deformed images are shown for different lesion filling methods.
    Red hollow arrows indicate artifacts in the filled regions, and red solid arrows indicate structural differences observed after registration.
    (b)~Deformed brain segmentations are compared to the target segmentation, with red solid arrows highlighting alignment patterns around ventricular regions.
	}
	\label{fig:figure4}
\end{figure*}

We compared MSRepaint lesion filling against FSL-filling~(v5.0\footnote{\url{https://process.innovation.ox.ac.uk/software/p/9564/fslv5/1}})~\citep{battaglini2012evaluating}, NiftySeg-filling~(commit \texttt{16cf563}\footnote{\url{https://github.com/KCL-BMEIS/NiftySeg}})~\citep{prados2016multi}, and FastSurfer-LIT~(commit \texttt{2d3652c}\footnote{\url{https://github.com/Deep-MI/LIT}}
)~\citep{pollak2025fastsurfer} using their publicly available code or installation packages.

\subsubsection{Lesion Filling Accuracy on BrainWeb Data}
\label{subsubsec:lesion_filling_brainweb}

To assess lesion-filling accuracy under controlled imaging conditions, we used the BrainWeb dataset (~\cite{cocosco1997brainweb}, Dataset~\#1 in Table~\ref{tab:dataset_summary}). 
This dataset provides a synthetic T1w brain image with severe lesions and a corresponding lesion-free reference image. 
We simulated different lesion-to-NAWM contrasts by applying gamma transformations ($\gamma = 1.0$, $1.25$, and $1.5$), enabling quantitative evaluation of filling performance across varying contrast levels.

Figure~\ref{fig:figure3} shows a qualitative and quantitative comparison of different lesion-filling methods.
Root mean squared error (RMSE) was computed within the lesion mask and normalized by the mean intensity of surrounding NAWM, providing a localized measure of inpainting accuracy.
Results show that across all $\gamma$ values, FSL-filling consistently produces the highest RMSEs compared to the other three methods and introduces visible spike-like artifacts~(red arrows).
NiftySeg-filling shows moderate RMSEs but still exhibits some artifact patterns~(red arrows), especially as $\gamma$ increases. 
In contrast, FastSurfer-LIT and MSRepaint-filling produce smooth, artifact-free results and consistently yield the lowest and comparable RMSEs, outperforming FSL-filling and NiftySeg-filling across all $\gamma$ values.
The computational times of the methods are approximately 2 minutes on CPU (Intel Xeon CPU E5-2620 v4) for FSL-filling, 5 minutes on CPU for NiftySeg-filling, 60 minutes on GPU (NVIDIA Quadro RTX 5000) for FastSurfer-LIT, and 3 minutes on GPU for MSRepaint-filling.

\subsubsection{Lesion Filling Effects on Registration-based Brain Volume Measurements}
\label{subsubsec:lesion_filling_brain_vol}

To evaluate the impact of lesion filling on registration-based brain volume measurements, we used the Private Real Lesions dataset (Dataset~\#2 in Table~\ref{tab:dataset_summary}). 
This dataset consists of 15 T1w images from PwMS, where lesion masks were generated using UNISELF~\citep{zhang2024isbi, zhang2025uniself}. 
We trained both VoxelMorph~\citep{balakrishnan2019voxelmorph} and Encoder-Only Image Registration~(EOIR)~\citep{chen2025encoder} models, the latter being a state-of-the-art registration model that ranked second in the 2024 LUMIR challenge~\citep{chen2025beyond}.  
The trained models were then used to estimate deformation fields between each lesion-filled T1w source and a separate lesion-free target, and these fields were applied to warp both the filled T1w image and its corresponding brain mask into the target space.  

\paragraph{Qualitative comparison} 
Figure~\ref{fig:figure4} shows a visual comparison of the results from different lesion-filling methods using VoxelMorph registration.
Figure~\ref{fig:figure4}a shows that FSL-filling introduces noise-like artifacts in the lesion-filled regions~(red hollow arrows), which result in noticeable misalignment of the ventricular structures~(red solid arrows) after registration.
While NiftySeg-filling reduces these artifacts in the lesion-filled regions~(red hollow arrows), it still leads to moderate structural misalignment~(red solid arrows).
In contrast, FastSurfer-LIT and MSRepaint-filling produce artifact-free lesion filling~(red hollow arrows), resulting in more anatomically consistent deformed images following registration~(red solid arrows).
Figure~\ref{fig:figure4}b shows the brain segmentation comparison between the target image and the deformed source images obtained using different lesion-filling methods.
Both FSL-filling and NiftySeg-filling result in noticeable misalignments, particularly around the ventricular regions~(red solid arrows).
In contrast, FastSurfer-LIT and MSRepaint-filling yield deformed brain masks that align closer with the target than the other two methods.

\begin{figure}[!t]
	\centering
    \includegraphics[width=1.0\columnwidth]{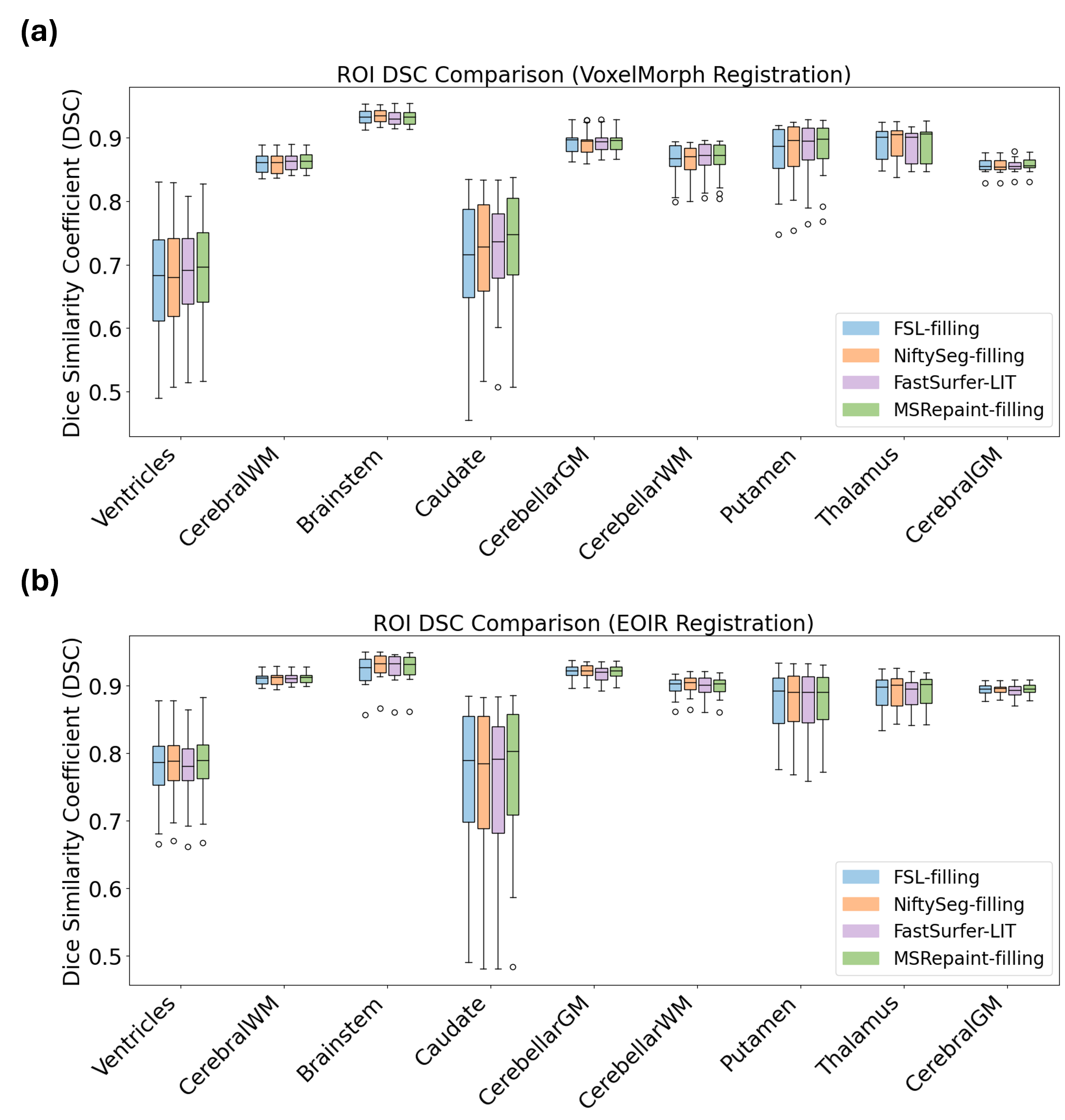}
	\caption{Region-wise Dice similarity coefficient (DSC) comparisons across 15 test subjects for registration-based brain subregion volume measurements following different lesion filling methods. 
    (a) VoxelMorph registration. 
    (b) EOIR registration. 
    No statistically significant differences were observed between MSRepaint-filling and the other methods (Wilcoxon signed-rank test, $p \geq 0.05$).
	}
	\label{fig:figure5}
\end{figure}

\paragraph{Quantitative Comparison}
Figure~\ref{fig:figure5} shows region-wise Dice similarity coefficient (DSC) comparisons across 15 test subjects for both registration models. 
With VoxelMorph (Fig.~\ref{fig:figure5}a), FastSurfer-LIT and MSRepaint-filling achieve consistently higher DSCs than FSL- and NiftySeg-filling in several lesion-sensitive regions, including the ventricles, caudate, and putamen. 
With EOIR (Fig.~\ref{fig:figure5}b), the relative improvements are less pronounced, though both FastSurfer-LIT and MSRepaint-filling remain comparable to, or slightly better than, FSL- and NiftySeg-filling. 
Across other regions, such as the brainstem, cerebellum, and cortical GM/WM, all four filling methods yield similar DSCs regardless of the registration model.

\subsubsection{Lesion Filling Effects on Deformation Field Accuracy in Deformable Registration}
\label{subsubsec:lesion_filling_deform}
\begin{figure}[!t]
	\centering
    \includegraphics[width=1.0\columnwidth]{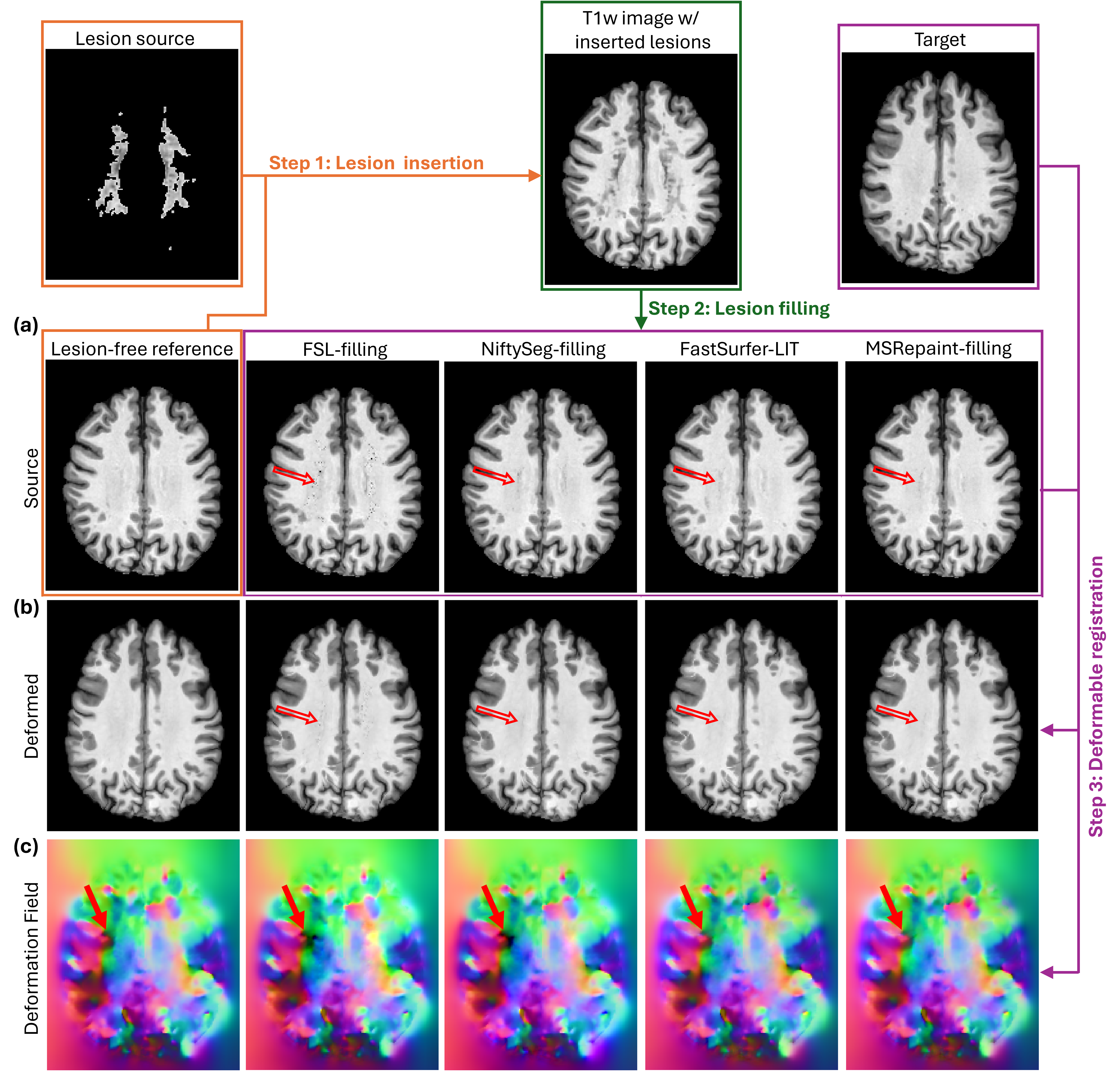}
	\caption{Evaluation of lesion-filling methods on deformation fields using EOIR registration. 
    (a)~Synthetic lesions were inserted into a lesion-free T1w image (Step 1), filled with different methods (Step 2). 
    (b)~Each filled image was registered to a lesion-free target, producing deformed images (Step 3). 
    (c)~Resulting deformation fields were compared with the reference from the original lesion-free image.
    Red hollow arrows indicate filling artifacts; red solid arrows indicate deformation field differences around lesion regions.
    RGB channels in (c)~represent displacement (R: $x$, G: $y$, B: $z$).
	}
	\label{fig:figure6}
\end{figure}

To evaluate the effect of lesion filling on deformation field accuracy in deformable registration, we used the Private Synthetic Lesions dataset (Dataset~\#3 in Table~\ref{tab:dataset_summary}). 
This dataset consists of 10 T1w images from healthy subjects, where lesions from separate MS subjects were inserted to simulate pathology. 
This design enables controlled experiments in which the ground-truth deformation field can be obtained by registering the original lesion-free image to a target, providing a reference for assessing how different lesion-filling methods influence registration accuracy.

\paragraph{Qualitative Comparison}
Figure~\ref{fig:figure6} shows a visual comparison of lesion-filling methods, where synthetic lesions were inserted into a lesion-free T1w image (Step 1), filled using different methods (Step 2), and registered to a lesion-free target with EOIR (Step 3), while the original lesion-free image was also registered to provide the reference deformation field.
In Fig.~\ref{fig:figure6}a, artifacts present in the lesion-filled source images using FSL-filling and NiftySeg-filling are substantially reduced in FastSurfer-LIT and MSRepaint-filling~(red hollow arrows).
These artifacts persist in the deformed outputs shown in Fig.~\ref{fig:figure6}b for both FSL-filling and NiftySeg-filling, but are absent in FastSurfer-LIT and MSRepaint-filling.
In Fig.~\ref{fig:figure6}c, FSL-filling and NiftySeg-filling produce deformation fields with exaggerated displacements around lesion regions~(red solid arrows), which are not observed in the reference deformation field.
In contrast, FastSurfer-LIT and MSRepaint-filling yield smoother deformation patterns that closely resemble the reference field~(red solid arrows).

\paragraph{Quantitative Comparison}

\begin{figure}[!t]
	\centering
    \includegraphics[width=1.0\columnwidth]{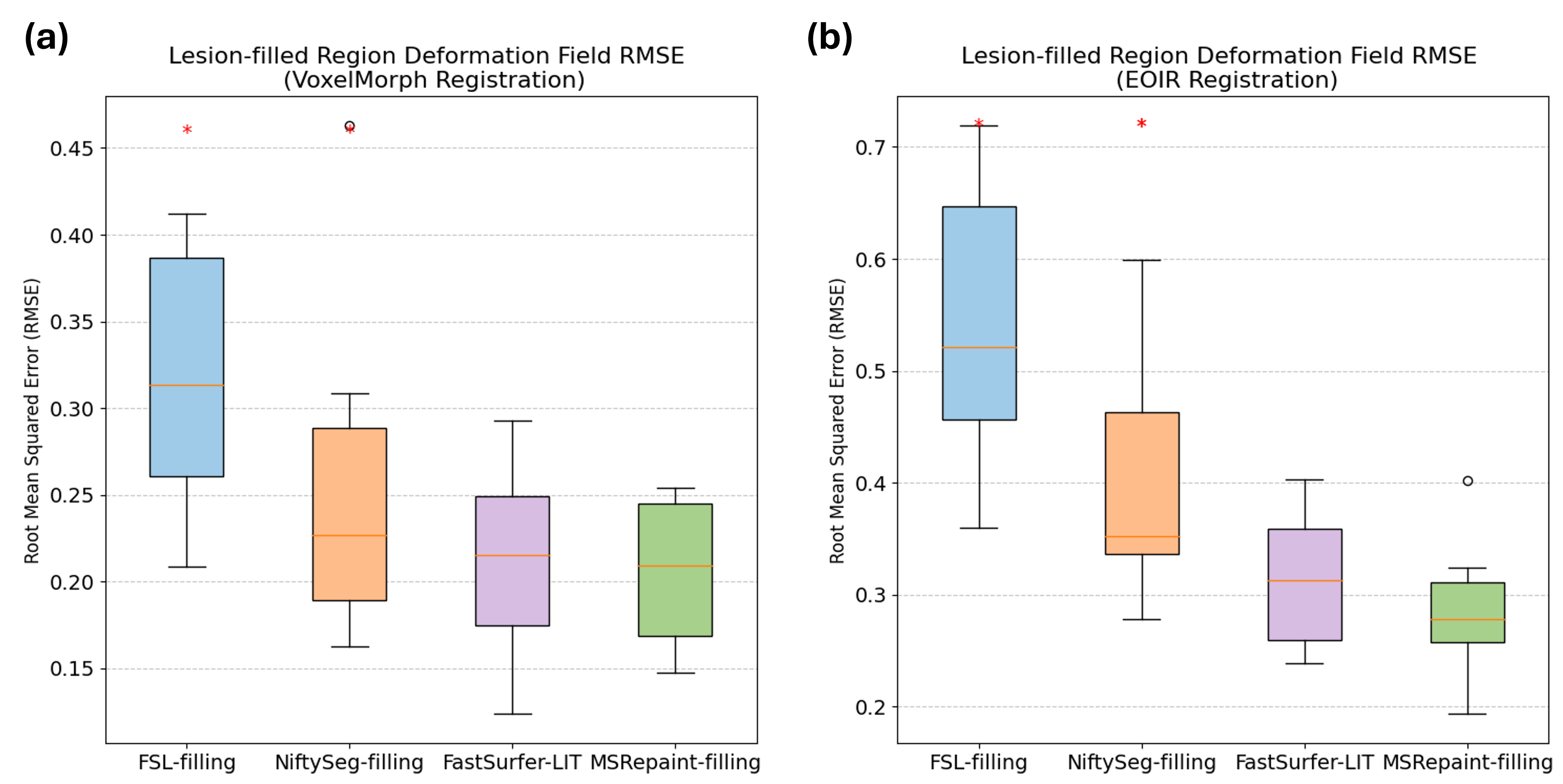}
	\caption{Deformation field RMSE within lesion regions relative to the reference across 10 test subjects following different lesion filling methods.
    (a) VoxelMorph registration.
    (b) EOIR registration.
    Red star: statistically significant difference between MSRepaint-filling and the other methods (Wilcoxon signed-rank test, $p < 0.05$).
	}
	\label{fig:figure7}
\end{figure}

Figure~\ref{fig:figure7} shows the deformation field RMSE within lesion regions relative to the reference field across 10 test subjects, with results from VoxelMorph (Fig.~\ref{fig:figure7}a) and EOIR (Fig.~\ref{fig:figure7}b).
In both settings, MSRepaint-filling yields significantly lower RMSEs than FSL-filling and NiftySeg-filling (Wilcoxon signed-rank test, $p < 0.05$).
Its performance also exhibits lower RMSEs than FastSurfer-LIT, although the difference is not statistically significant.

\subsection{Lesion Synthesis Evaluation}

We compared MSRepaint lesion synthesis against CarveMix~\citep{zhang2023carvemix} using its publicly available code~(commit \texttt{e5942cd}\footnote{\url{https://github.com/ZhangxinruBIT/CarveMix}}).
We did not include other lesion synthesis methods in our comparison due to incomplete documentation in their public repositories~\citep{salem2019multiple}, lack of support for multicontrast and 3D MRI~\citep{basaran2022subject, mathur2025long}, or infeasible GPU memory requirements for high-resolution brain MRI~\citep{zhang2024lefusion}.

To evaluate the effectiveness of lesion synthesis, we trained UNISELF~\citep{zhang2024towards, zhang2025uniself}, a state-of-the-art MS lesion segmentation method with high in-domain accuracy and strong out-of-domain generalization, on various synthetic training datasets and evaluated its performance on the 2016 MICCAI Challenge~\citep{commowick2018objective} (Dataset \#6 in Table~\ref{tab:dataset_summary}) and UMCL~\citep{lesjak2018novel} (Dataset \#7 in Table~\ref{tab:dataset_summary}) datasets.
A UNISELF model trained on the original ISBI challenge~\citep{carass2017longitudinal} training dataset (Dataset \#5 in Table~\ref{tab:dataset_summary}) was used as the baseline.
Three synthetic training datasets were generated for UNISELF training, described as follows:
\begin{enumerate}
    \item \textbf{CarveMix on ISBI}: Generated by applying CarveMix to the ISBI training dataset, creating synthetic lesion masks and the corresponding mixed-up images.
    
    \item \textbf{MSRepaint on ISBI with CarveMix lesion masks}: Generated by replacing the image mixing step in CarveMix with MSRepaint synthesis, using the CarveMix-augmented lesion masks as conditional inputs.
    
    \item \textbf{MSRepaint with lesion dictionaries}: Generated using the lesion dictionary-based synthesis pipeline described in Section~\ref{subsec:lesion_dictionary_and_synthesis}, with lesion dictionaries constructed using Dataset \#4 in Table~\ref{tab:dataset_summary}.

\end{enumerate}

For synthetic datasets 1 and 2, the five anatomical spaces correspond to the five PwMS in the original ISBI training set. 
For synthetic dataset 3, we also included five anatomical spaces using our lesion dictionary pipeline (Section~\ref{subsec:lesion_dictionary_and_synthesis}), each anchored to a different healthy subject, consistent with the number of anatomical spaces in the original ISBI training set. 
In all three synthetic datasets, 200 synthetic images were generated per anatomical space, yielding 1,000 multicontrast images per dataset.
This design preserved the ISBI structure of five subject spaces for training while expanding the dataset size in terms of lesion diversity to evaluate lesion synthesis methods.

\subsubsection{CarveMix vs. MSRepaint Synthesis with CarveMix-augmented Lesion Masks}
\label{subsubsec:carvemix_msrepaint_synthesis}

\begin{figure}[!t]
	\centering
    \includegraphics[width=1.0\columnwidth]{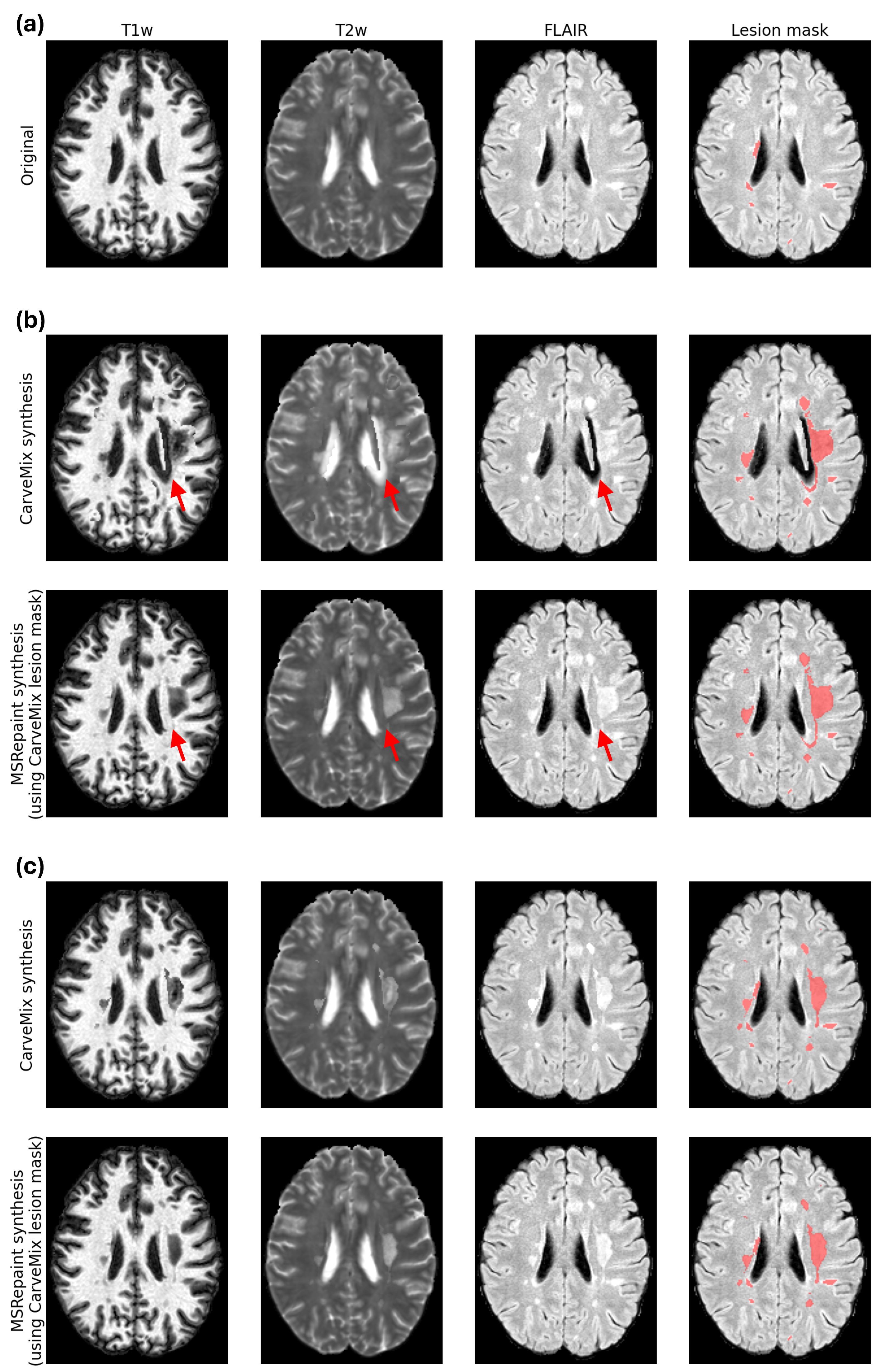}
	\caption{Comparison of lesion synthesis results on a representative ISBI subject using CarveMix and MSRepaint, both generated from CarveMix-augmented lesion masks. 
    (a)~Original multicontrast MRI and lesion mask. 
    (b)~First example of synthetic images generated by CarveMix~(first row) and MSRepaint~(second row) with the corresponding CarveMix-augmented lesion mask (red arrows indicate lesion boundaries).
    (c)~Second example of synthetic images generated by CarveMix~(first row) and MSRepaint~(second row) with the corresponding CarveMix-augmented lesion mask.
	}
	\label{fig:figure8}
\end{figure}

Figure~\ref{fig:figure8} compares lesion synthesis results on a representative ISBI subject using CarveMix and MSRepaint, both generated from CarveMix-augmented lesion masks.
Figure~\ref{fig:figure8}a shows the original multicontrast MRI and the corresponding lesion mask.
Figure~\ref{fig:figure8}b shows one comparison of synthetic images from CarveMix (first row) and MSRepaint (second row). In the CarveMix output, lesions are embedded but accompanied by visible artifacts, particularly intensity discontinuities near lesion boundaries (red arrows). 
In contrast, MSRepaint produces more anatomically plausible results, with smoother lesion integration across contrasts and no artificial discontinuities.
Figure~\ref{fig:figure8}c shows another comparison example.
In this case, while the CarveMix output (first row) shows no obvious discontinuities, the MSRepaint output (second row) still demonstrates more realistic tissue transitions from synthetic lesions to NAWM.

\begin{figure*}[!t]
	\centering
    \includegraphics[width=2.0\columnwidth]{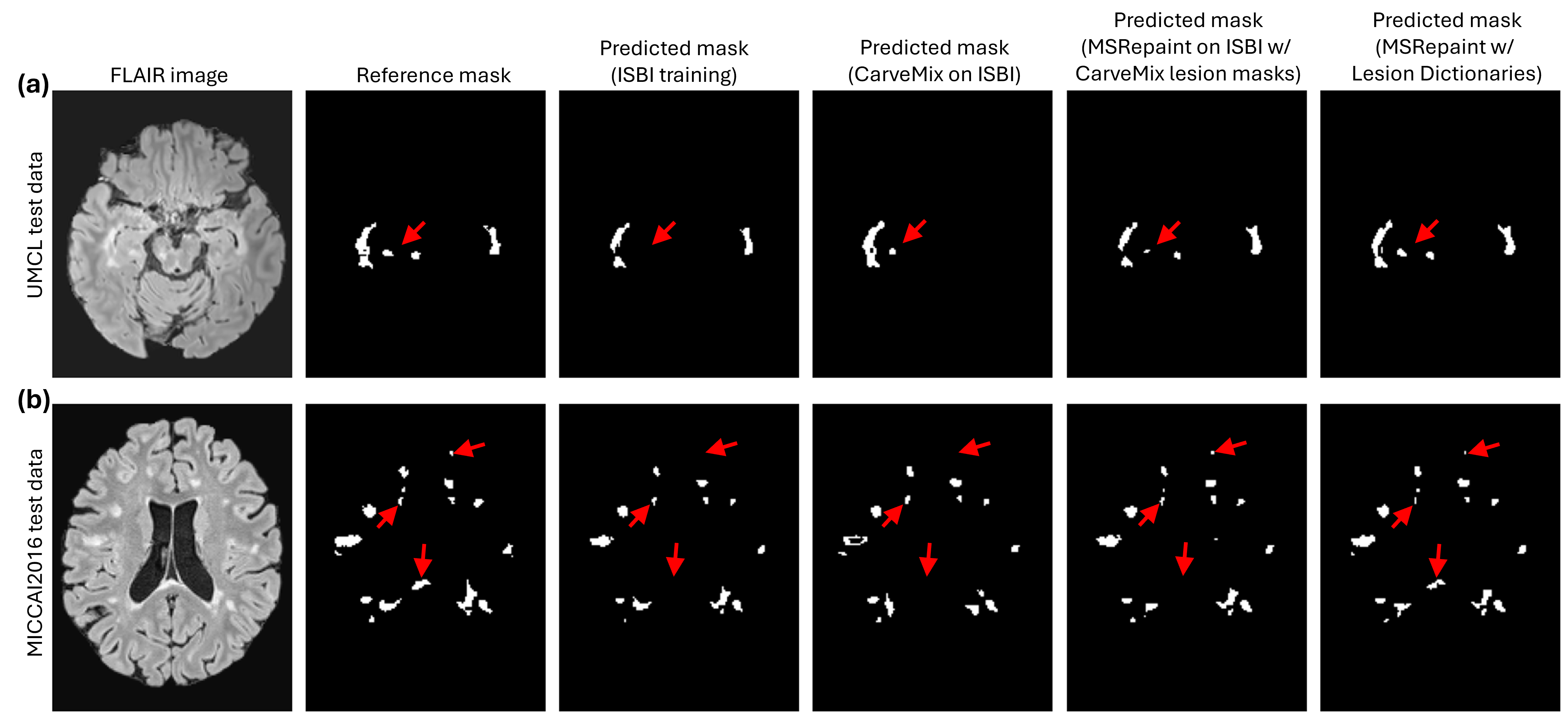}
	\caption{Comparison of MS lesion segmentation results on two out-of-domain test datasets, UMCL~(a) and MICCAI2016~(b), using the UNISELF model trained on different datasets.
    Each row shows the input FLAIR image, reference lesion mask, and predicted lesion masks from models trained on: (1) original ISBI training data, (2) CarveMix-synthesized data based on ISBI, (3) MSRepaint-synthesized data based on ISBI using CarveMix lesion masks, and (4) MSRepaint-synthesized data using lesion dictionaries~(Section~\ref{subsec:lesion_dictionary_and_synthesis}). 
    Red solid arrows indicate lesions for visual comparison across methods.
	}
	\label{fig:figure10}
\end{figure*}

\subsubsection{Qualitative Comparison of Lesion Segmentation}
\label{subsubsec:lesion_seg_qualitative}

Figure~\ref{fig:figure10} shows qualitative comparisons of MS lesion segmentation on two out-of-domain test datasets, UMCL~(a) and MICCAI2016~(b). 
Each row shows from left to right: the input FLAIR image, the reference lesion mask, and predicted lesion masks from UNISELF models trained on: (1) the original ISBI training dataset, (2) a CarveMix-synthesized dataset based on ISBI, (3) an MSRepaint-synthesized dataset based on ISBI using the same CarveMix-generated lesion masks, and (4) an MSRepaint-synthesized dataset using lesion dictionaries. 
Red solid arrows highlight lesions in the reference masks that are missed or incompletely segmented by one or more models. 
The model trained on the ISBI dataset misses several lesions, particularly in challenging regions such as the brainstem (Fig.~\ref{fig:figure10}a) and periventricular areas~(Fig.~\ref{fig:figure10}b). 
The CarveMix-based model shows similar or worse limitations, with more missed detections and under-segmentations in multiple locations. 
In contrast, both MSRepaint-based models, especially the one trained with lesion dictionaries, demonstrate improved lesion coverage and more complete segmentation, including better detection of lesions in difficult anatomical regions.

\subsubsection{Quantitative Comparison of Lesion Segmentation}
\label{subsubsec:lesion_seg_quantitative}

\begin{figure}[!t]
	\centering
    \includegraphics[width=1.0\columnwidth]{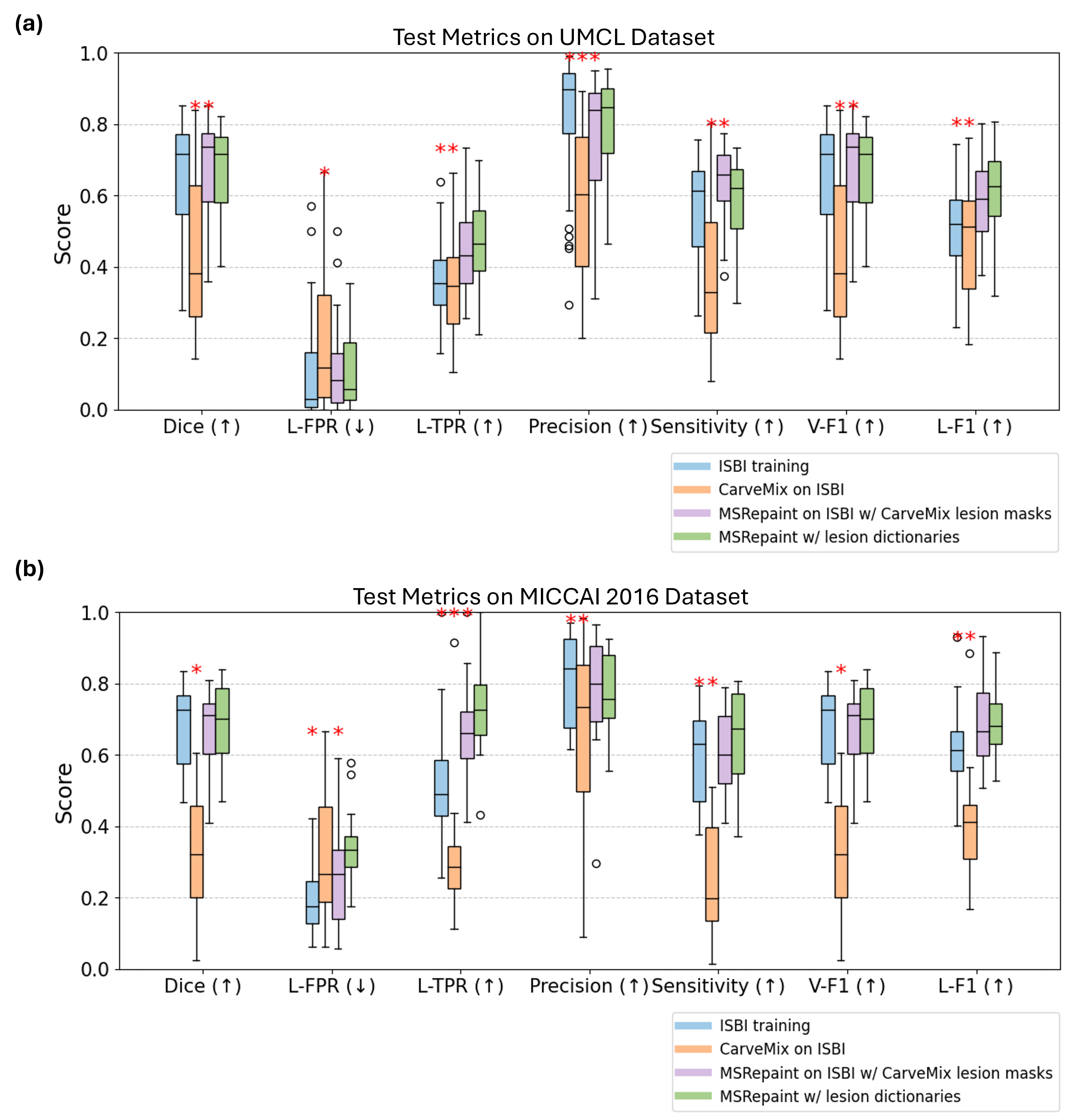}
	\caption{Quantitative comparison of MS lesion segmentation performance on two test datasets, UMCL~(a) and MICCAI2016~(b), using the UNISELF model trained on different datasets. 
    Metrics include both voxel-level (Dice, Precision, Sensitivity, V-F1) and lesion-level (L-FPR, L-TPR, L-F1) evaluations.
    Red star: statistically significant difference between the MSRepaint with lesion dictionaries (last boxplot) and each of the other three datasets (Wilcoxon signed-rank test, $p < 0.05$).
	}
	\label{fig:figure11}
\end{figure}

Figure~\ref{fig:figure11} shows quantitative comparisons of MS lesion segmentation performance across the two test datasets using different training data. 
For both test datasets, models trained with the two MSRepaint-synthesized data consistently achieved higher lesion-wise F1 scores (L-F1) than those trained with the original ISBI data or CarveMix-augmented ISBI data, indicating improved lesion-level detection performance.
These improvements were statistically significant, as confirmed by Wilcoxon signed-rank tests (last subplot in Figs.~\ref{fig:figure11}a and b).
Moreover, MSRepaint synthesis with lesion dictionaries derived from large-scale private data achieved slightly higher L-F1 compared to MSRepaint synthesis using ISBI data with CarveMix lesion masks.
In contrast, the CarveMix-augmented ISBI data led to lower performance than the original ISBI data across most metrics, including Dice, L-F1, and voxel-wise F1 (V-F1), with increased variability.
In addition to lesion-level improvements, models trained on MSRepaint-synthesized data achieved comparable voxel-level segmentation accuracy to the ISBI-trained model, as measured by Dice and V-F1.

\subsection{Lesion Evolution Simulation Evaluation}
\label{subsec:lesion_evo}

\begin{figure}[!t]
	\centering
    \includegraphics[width=1.0\columnwidth]{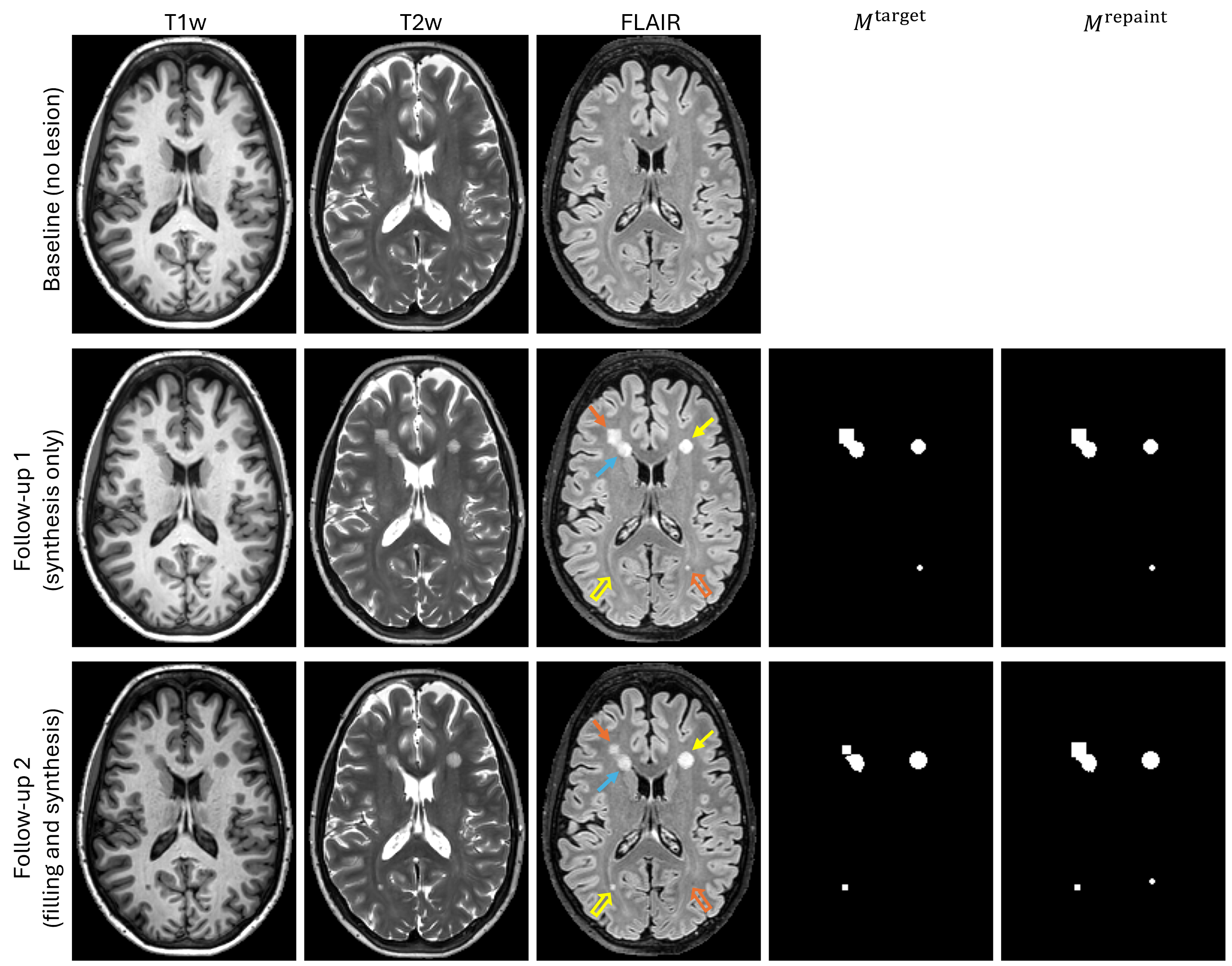}
	\caption{Example of longitudinal lesion evolution simulation using MSRepaint.
        The first row shows baseline images without lesions.
        The second row shows lesion synthesis on a follow-up scan using $M^{\text{target}}$ with both cubic/spherical shapes and a realistic-appearing lesion (teal arrow), with $M^{\text{repaint}} = M^{\text{target}}$.
        The third row shows longitudinal simulation of lesion evolution on the second follow-up scan, including growth (yellow solid arrow), shrinkage (orange solid arrow), appearance (yellow hollow arrow), disappearance (orange hollow arrow), and evolution of a realistic lesion (teal arrow), with $M^{\text{repaint}} \supseteq M^{\text{target}}$.
}

	\label{fig:Figure12}
\end{figure}

To evaluate MSRepaint’s ability to simulate longitudinal lesion evolution under full mask control, we used the Private Longitudinal Data (Dataset~\#8 in Table~\ref{tab:dataset_summary}). 
This dataset consists of multicontrast MRI from a single healthy subject, with spatial guidance masks manually designed to simulate lesion changes across two follow-up timepoints. 
Starting from a baseline lesion-free scan, we applied MSRepaint with these masks to synthesize temporal lesion dynamics, including appearance, disappearance, growth, shrinkage, and realistic lesion evolution.

Figure~\ref{fig:Figure12} visualizes this longitudinal simulation.  
Each row shows multicontrast images (T1w, T2w, and FLAIR) at different timepoints, along with the corresponding masks $M^{\text{target}}$ and $M^{\text{repaint}}$ used by MSRepaint.  
The first row shows baseline images without lesions.  
The second row shows lesion synthesis on a follow-up scan, where new synthetic lesions are introduced based on the spatial guidance mask $M^{\text{target}}$ and the repaint mask ($M^{\text{repaint}} = M^{\text{target}}$).  
Here, both non-physiological cubic/spherical shapes (orange and yellow arrows) and a realistic-appearing lesion (teal arrow) are generated to highlight that MSRepaint can faithfully follow arbitrary mask inputs while also producing anatomically plausible lesion patterns.
The third row shows diverse lesion evolution patterns on the second follow-up scan, including lesion growth (yellow solid arrow), shrinkage (orange solid arrow), appearance (yellow hollow arrow), disappearance (orange hollow arrow), and evolution of a realistic lesion (teal arrow).
In this case, $M^{\text{target}}$ specifies the lesion configuration, while $M^{\text{repaint}} \supseteq M^{\text{target}}$ denotes the regions where both lesion synthesis and filling are applied through repainting.
For the growth and shrinkage cases, the cubic and spherical lesions preserve their geometry, while the realistic lesion maintains a natural morphology, underscoring MSRepaint’s controllability in handling both arbitrary and anatomically plausible lesion shapes.

\section{Discussion}
\label{s:discussion}
This work introduces and evaluates MSRepaint, a unified framework for bidirectional MS lesion filling and synthesis in brain MRI.
MSRepaint consistently outperforms existing methods in both lesion filling and synthesis across multiple experimental settings, yielding notable downstream improvements in brain segmentation, image registration, and lesion segmentation.
MSRepaint also enables simulating lesion evolution with a unified model, a capability never achieved by any related prior work.

\subsection{Lesion Filling: Improved Image Fidelity and Downstream Robustness}

Compared to FSL-filling and NiftySeg-filling, MSRepaint-filling consistently produces smoother, artifact-free lesion-filled images with lower voxel-wise reconstruction errors under varying lesion-to-NAWM contrast ratios (Fig.~\ref{fig:figure3}).
These improvements are attributed to MSRepaint’s generative modeling capability, which effectively captures contextual anatomical features to enable more plausible inpainting.
Importantly, although FastSurfer-LIT achieves similarly low MSEs, MSRepaint performs inference approximately 20 times faster (60 vs. 3 minutes on GPU) and requires far less training data (1,315 vs 76 volumes), demonstrating superior efficiency and effectiveness.

Beyond image quality, MSRepaint contributes to more accurate downstream brain morphometry.
When integrated with deformable registration for brain volume measurement, MSRepaint-filling reduces the impact of lesion artifacts on regional volume estimates, leading to improved alignment of ventricles (Fig.~\ref{fig:figure4}) and higher Dice scores for anatomical segmentation of lesion-sensitive structures (e.g., ventricles, caudate, putamen) (Fig.~\ref{fig:figure5}).
These improvements hold across different registration models (VoxelMorph and EOIR), demonstrating consistent gains in anatomical alignment and segmentation accuracy regardless of the registration method used.
While FastSurfer-LIT achieves comparable anatomical accuracy, its significantly longer runtime may limit scalability in clinical or limited resource settings.

MSRepaint also improves deformable registration accuracy. 
It reduces lesion-induced artifacts (Fig.\ref{fig:figure6}) and yields significantly lower deformation field errors than FSL-filling and NiftySeg-filling (Fig.\ref{fig:figure7}).
It matches or slightly outperforms FastSurfer-LIT in accuracy while providing much faster inference with far less training data, underscoring its robustness and efficiency for deformable registration with lesions.
This capability is crucial for longitudinal MS lesion tracking in large cohorts, as shown in our recent work~\citep{zhang2025aultra}, where MSRepaint-preprocessed intra-subject registration accounts for brain atrophy and enables unique lesion identification~\citep{rivas2025unique} longitudinally to support lesion-wise morphometric analysis.

\subsection{Lesion Synthesis: Data Augmentation with Anatomical Realism}

MSRepaint also demonstrates superior performance over existing methods in generating synthetic data for training lesion segmentation models.
Compared to CarveMix, which relies on direct image mixing and often introduces intensity discontinuities, MSRepaint produces anatomically realistic lesion appearances with seamless tissue transitions (Fig.~\ref{fig:figure8}).
This improved visual plausibility translates into better segmentation generalization on out-of-domain datasets~(Figs.~\ref{fig:figure10} and~\ref{fig:figure11}).
UNISELF segmentation models trained with MSRepaint-synthesized data, whether using CarveMix masks or dictionary-based masks, consistently achieve higher L-F1 than UNISELF models trained with either the original ISBI dataset or CarveMix-augmented ISBI data (Fig.~\ref{fig:figure11}).
These improvements are statistically significant and consistent across both UMCL and MICCAI2016 test sets.
Notably, MSRepaint-synthesized datasets using lesion dictionaries yield the best and most stable L-F1, supporting the value of population-derived lesion distribution in generating diverse and challenging training examples.
Furthermore, these improvements occur without compromising voxel-level segmentation accuracy, as evidenced by comparable Dice and V-F1 scores.

Compared to other deep learning-based synthesis approaches, MSRepaint offers an efficient and anatomically consistent framework for MS lesion synthesis.
CNN-based models~\citep{salem2019multiple} typically rely on supervised training with pseudo-healthy inputs and exhibit limited generative diversity, restricting their ability to simulate the wide variability in lesion appearance observed in MS.
GAN-based approaches~\citep{basaran2022subject} introduce stochasticity and realism but suffer from known issues such as mode collapse, training instability, and limited spatial control. Diffusion-based models~\citep{zhang2024lefusion, mathur2025long} offer improved fidelity and stability, yet often require large-scale training data and high computational resources.
MSRepaint addresses these limitations by supporting spatial lesion mask guidance with limited training data, multicontrast inputs with dropout robustness, efficient inference via multi-view DDIM, and large-scale data generation through lesion dictionaries.

\subsection{A Unified Framework for Bidirectional Lesion Manipulation}

A central innovation of MSRepaint is its unified treatment of lesion filling and synthesis through spatial guidance masks. 
As demonstrated in the longitudinal lesion evolution simulation (Fig.~\ref{fig:Figure12}), the framework can model key patterns of lesion dynamics, including appearance, disappearance, growth, and shrinkage, within a single generative model. 
Importantly, these simulations highlight the full controllability of MSRepaint: lesion shape, size, location, and temporal evolution are entirely controlled by the input masks. 
This was illustrated using deliberately cubic and spherical masks that were never encountered during training; the model nevertheless reproduced them faithfully across timepoints, underscoring its capacity to follow arbitrary guidance rather than overfitting to realistic lesion morphology. 

Such controllability has important implications. 
First, it enables controlled data augmentation for temporal modeling, supporting applications such as clinical trial simulation and disease progression analysis. 
Second. by generating time-resolved lesion trajectories with explicit control over spatial and temporal properties, MSRepaint can help benchmark lesion tracking algorithms and evaluate their sensitivity to subtle changes over time.

\subsection{Limitations and Future Work}

Despite its strengths, several limitations merit discussion.
First, although MSRepaint supports spatial lesion mask guidance and multicontrast inputs, it does not explicitly model lesion temporal dynamics.
Developing spatiotemporal lesion evolution models and integrating them into MSRepaint would enable more comprehensive simulation of disease activity.
Second, while MSRepaint filling and synthesis improve downstream analysis and image realism, clinical validation of their impact, such as on diagnosis, prognosis, or treatment planning, remains limited.
Future work should apply MSRepaint to large-scale clinical trial MRI datasets to assess its potential for improving automated processing pipelines and enabling more robust MS imaging analysis.
Third, although MSRepaint has been developed and evaluated using structural brain MRI data from MS cohorts, its applicability to other neurological conditions, such as stroke, brain tumors, or age-related CSF changes, or to other imaging modalities, such as diffusion-weighted imaging, remains unexplored.
Future work should extend MSRepaint to additional disease domains and imaging modalities to broaden its applicability across diverse neuroimaging contexts.

\section{Conclusion}
\label{s:conclusion}
Overall, MSRepaint provides a robust, efficient, and versatile solution for MS lesion manipulation in brain MRI.
It achieves superior performance in both lesion filling and synthesis, improves downstream tasks including brain segmentation, image registration, and lesion segmentation, and offers a flexible framework for simulating lesion evolution.
These capabilities make MSRepaint a valuable tool for neuroimaging applications in both research and clinical settings. 

\section*{Acknowledgments}
This work was partially supported by NSF GRFP DGE-1746891~(S.~W.~Remedios); CDMRP W81XWH2010912~(PI: J.~L.~Prince) and HT94252510716~(PI: J.~Zhang); NMSS RG-1507-05243 (PI: D.~L.~Pham); NIH R01-NS082347 (PIs: S.~Saidha and P.~A.~Calabresi) and U01-NS111678 (PI: P.~A.~Calabresi); NIH R01-CA253923~(PI: B.~A.~Landman) and R01-CA275015~(PI: B.~A.~Landman), which also provided support for L. Zuo; and PCORI grant MS-1610-37115~(PIs: S.~D.~Newsome and E.~M.~Mowry), which also provided support for J. Zhang, L. Zuo, S. W. Remedios, and A. Carass. 
Y. Liu was supported by the NIH grants R01-HL169944, U24-AG074855, and R01-MH121620.
The statements in this publication are solely the responsibility of the authors and do not necessarily represent the views of PCORI, its Board of Governors or Methodology Committee.

\bibliographystyle{model2-names}
\biboptions{authoryear}
\bibliography{refs}

\clearpage

\end{document}